\documentclass[a4paper,fleqn,usenatbib]{mnras}

\usepackage{txfonts}

\usepackage[T1]{fontenc}
\usepackage{ae,aecompl}

    \hypersetup{pdfauthor={Timothy. A. Davis},
               pdftitle={Spatially resolved variations of the IMF mass normalisation in early-type galaxies as probed by molecular gas kinematics},
               pdfkeywords={galaxies: elliptical and lenticular, cD -- ISM: molecules -- galaxies: ISM -- galaxies: evolution -- galaxies: kinematics and dynamics},
               bookmarksnumbered=true}

\usepackage{amssymb}	
\usepackage{graphicx,color,afterpage}
\usepackage{times,rotating}
\usepackage{float}
 \volume{{\rm in press}}

\def\hi{\mbox{H\sc{i}}}

\def\atlas{{{ATLAS}}$^{\rm 3D}$}

\def\kms{km s$^{-1}$}

\def\arcsec{$^{\prime \prime}$}
\definecolor{Mygrey}{gray}{0.75}

\newcommand{\ltsimeq}{\raisebox{-0.6ex}{$\,\stackrel{\raisebox{-.2ex}{$\textstyle <$}}{\sim}\,$}}

\newcommand{\farc}{\mbox{\ensuremath{.\!\!^{\prime\prime}}}}

\mathchardef\mhyphen="2D

\usepackage[compact]{titlesec}
\titlespacing{\section}{0pt}{*2}{*1}

\title[Spatially resolved IMF gradients in ETGs]{Spatially resolved variations of the IMF mass normalisation in early-type galaxies as probed by molecular gas kinematics} 
\author[Timothy A. Davis et al.]{\parbox{\textwidth}{Timothy A. Davis$^{1}$\thanks{E-mail: \texttt{DavisT@cardiff.ac.uk}} and Richard M. McDermid$^{2,3}$}
\vspace{0.4cm}\\
\parbox{\textwidth}{$^{1}$School of Physics \&\ Astronomy, Cardiff University, Queens Buildings, The Parade, Cardiff, CF24 3AA, UK\\
$^{2}$Department of Physics and Astronomy, Macquarie University, Sydney, NSW 2109, Australia\\
$^{3}$Australian Astronomical Observatory, PO Box 915, Sydney, NSW 1670, Australia}}
\begin{document}
\date{Accepted 2016 September 12. Received 2016 September 12; in original form 2016 May 5}

\pagerange{\pageref{firstpage}--\pageref{lastpage}} \pubyear{2015}

\maketitle

\label{firstpage}

\begin{abstract}
We here present the first spatially-resolved study of the IMF in external galaxies derived using a dynamical tracer of the mass-to-light ratio.
We use the kinematics of relaxed molecular gas discs in seven early-type galaxies (ETGs) selected from the \atlas\ survey to dynamically determine mass-to-light ratio (M/L) gradients.
These M/L gradients are not very strong in the inner parts of these objects, and galaxies that do show variations are those with the highest specific star formation rates. 
Stellar population parameters derived from star formation histories are then used in order to estimate the stellar initial mass function function (IMF) mismatch parameter, and shed light on its variation within ETGs. 
Some of our target objects require a light IMF, otherwise their stellar population masses would be greater than their dynamical masses. In contrast, other systems seem to require heavier IMFs to explain their gas kinematics. Our analysis again confirms that IMF variation seems to be occurring within massive ETGs. We find good agreement between our IMF normalisations derived using molecular gas kinematics and those derived using other techniques.
Despite this, we do not see find any correlation between the IMF normalisation and galaxy dynamical properties or stellar population parameters, either locally or globally. In the future larger studies which use molecules as tracers of galaxy dynamics can be used to help us disentangle the root cause of IMF variation. 
\end{abstract}


\section{Introduction}

The stellar initial mass function function (IMF) is one of the most fundamental, and hotly debated, observational topics in astrophysics. Observations of stars within our own Milky Way suggest that the gravitational collapse of molecular clouds leads to star formation, and the birth of a population of stars whose masses can be well described by a single mass function \citep{1955ApJ...121..161S,2001MNRAS.322..231K,2003PASP..115..763C}. This mass function appears to be universal across the range of environments which we are able to probe within our own Galaxy \citep{2002Sci...295...82K,2010ARA&A..48..339B}.

In the early universe, and in other extragalactic environments, however, conditions can be very different than those found locally. 
Understanding if the IMF is universal in these places, or if it varies (and why) is crucial to allow interpretations of observations, impacting almost all areas of astrophysics. For instance calibrations that allow estimation of star formation rates, stellar mass loss return rates and even total stellar masses rely intimately on an assumed IMF \citep[e.g.][]{2016arXiv160305281C}. 

Recently, evidence for the non-universality of the stellar IMF of the most massive early-type galaxies (ETGs) has begun to mount. 
This evidence comes from three independent techniques which can probe the IMF in the
unresolved stellar populations of ETGs, based on the
modelling of stellar kinematics (e.g.~\citealt{2012Natur.484..485C,2013MNRAS.432.2496D,2013ApJ...765....8T}); utilising strong
gravitational lensing \citep[e.g.][]{2010ApJ...709.1195T,2010ApJ...721L.163A}; or via gravity-sensitive
spectroscopic features in galaxy spectra (e.g.~\citealt{2003MNRAS.339L..12C,2010Natur.468..940V,2012ApJ...760...71C,2013MNRAS.429L..15F}).

Stellar kinematic and lensing studies are sensitive to the stellar M/L, and the major
uncertainty in such studies relates to the contribution of the dark matter halo to the potential of galaxies. Spectroscopic constraints are mostly
sensitive to the ratio between giant and dwarf stars, and its primary uncertainty is degeneracy between the IMF parameters and those of the
underlying stellar populations, most notably the effect of variations
in the individual elemental abundances.

Studies utilising these techniques seem to agree that a variation in the IMF is occurring, with more massive ETGs having heavier IMFs. These studies disagree, however, on what the primary driving mechanism for such a variation is, with some studies favouring galaxy velocity dispersion \citep[e.g.][]{2012Natur.484..485C,2013MNRAS.433.3017L,2015MNRAS.446..493P}, others metallicity \citep[e.g.][]{2015ApJ...806L..31M} or (alpha-)element abundances \citep[e.g.][]{2012ApJ...760...71C}. Some authors have suggested that these studies lack internal consistency, with different analyses of the same objects finding different IMF slopes \citep{2014MNRAS.443L..69S}. The water has muddied further with the discovery that dwarf elliptical galaxies also seem to have non universal IMFs, spanning a similar range of giant galaxies, while having vastly different properties \citep{2015arXiv150908462T}.

In this work we introduce a new complementary technique to probe the IMF in galaxies. We use the kinematics of the cold molecular gas reservoirs in massive early-type galaxies to constrain their mass profiles. By combining these profiles with observations of the galaxies' stellar luminosity profile (and stellar population parameters) we are able to derive mass-to-light ratios, and constrain the IMF in a radially resolved manner. Studies of the radial variation of the IMF within individual objects are relatively new (see e.g. \citealt{2015arXiv150908250L,2015MNRAS.452..597Z}), but have significant diagnostic power to determine the astrophysics behind IMF variation. For instance \cite{2015MNRAS.447.1033M} find significant IMF gradients in two massive ETGs, while a lower mass object showed little variation. If confirmed, this could imply that the enhanced fraction of low mass stars causing IMF variation is only present in galaxy bulges which formed violently at high redshift.  

In Section \ref{sample} of this paper we present details of our sample selection and the properties of the target objects. Section \ref{newdata} details the observation parameters and reduction for the new data used in this work. In Section \ref{method} we present details of the method we use to constrain the IMF.  We then present our results in Section \ref{results}, and discuss them in Section \ref{discuss}, before concluding in Section \ref{conclude}.

 \begin{table*}
\caption{Properties of the ETGs included in this study}
\begin{tabular*}{0.85\textwidth}{@{\extracolsep{\fill}}l r r r r r r r r}
\hline
Name & Distance & M$_{Ks}$ & $\sigma_{e}$ & R$_e$ & R$_{\rm max}$/R$_e$ & log$_{10}$(M$_{\rm gas}$/M$_*$) &$\alpha_{\rm dyn}$ (C+12) \\ 
  & (Mpc) &  (mag) & (\kms) & (kpc) & & \\
 (1) & (2) & (3) & (4) & (5) & (6) & (7) & (8)\\
 \hline
 NGC0524 & 23.3 & -24.71 & 220 & 4.9 & 0.49 & -3.43 &  0.60\\
 NGC3607 & 22.2 & -24.74 & 206 & 4.1 & 0.21 & -2.84 &   0.72\\
 NGC3665 & 33.1 & -24.92 & 216 & 5.0 & 0.51 & -2.44 &   0.96\\
 NGC4429 & 16.5 & -24.32 & 177 & 3.3 & 0.52 & -2.78 &   0.92\\
 NGC4459 & 16.1 & -23.89 & 158 & 2.8 & 0.13 & -2.67 &   0.70\\
 NGC4526 & 16.4 & -24.62 & 208 & 3.5 & 0.35 & -2.65 &   0.94\\
 IC0719 & 29.4  & -22.70 & 128 & 1.8  & 1.11  & -2.29 & $^*$2.06  \\
        \hline
\end{tabular*}
\parbox[t]{0.85\textwidth}{{ \textit{Notes:}  Column 1 lists the name of each source. Column 2 to 5 are the distance, $Ks$-band absolute magnitude, velocity dispersion within one effective radius, and effective radius of each object. These are reproduced from \cite{2011MNRAS.413..813C} and \cite{2013MNRAS.432.1709C}. Column 6 contains the ratio of R$_{\rm max}$ (the radius at which the rotation profile becomes flat; these figures taken from \citealt{2014MNRAS.444.3427D}) to the effective radius R$_e$. Column 7 lists the gas fraction (molecular plus atomic) within the inner regions of these objects, as described in \cite{2014MNRAS.444.3427D}. The stellar mass used here is the dynamical mass derived from jeans modelling in \cite{2013MNRAS.432.1709C}. Column 8 contains the $\alpha_{\rm dyn}$ value derived by \cite{2012Natur.484..485C}. A star denotes values of $\alpha_{\rm dyn}$ considered unreliable by \cite{2012Natur.484..485C} due to the presence of strong population gradients.}}
\label{proptable}
\end{table*}

\section{Sample}
\label{sample}

In this work we require resolved maps of the molecular interstellar medium (ISM) in massive early-type galaxies. We thus selected objects from the \atlas\ survey \citep{2011MNRAS.413..813C}, which has provided the largest sample of interferometrically mapped ETGs to date \citep{2013MNRAS.432.1796A}. In addition, by selecting from this survey we have access to all the required ancillary data (including integral field spectroscopy and optical imaging) required to complete our analysis. 
We select ETGs from the \atlas\ survey which have been shown in \cite{2013MNRAS.432.1796A} and \cite{2013MNRAS.429..534D} to have regular, relaxed molecular gas distributions that rotate in the disc plane of their host \citep{2011MNRAS.417..882D}. We require that the molecular discs in these objects are well resolved, with at least 3 beams radially along their major axis. This leads to a sample of seven objects that we study in this work. These objects span a range of galaxy parameters, with velocity dispersions between 150 and 260 \kms\ \citep{2013MNRAS.432.1709C}, stellar metallicities of a 2/3 to 1.5 times solar, and mean mass weighted stellar ages of 7 to 14 Gyr \citep{2015MNRAS.448.3484M}. Other salient properties of these objects are described in Table \ref{proptable}.

For three of these objects we utilise the CO(1-0) interferometric data of \cite{2013MNRAS.432.1796A}, taken as part of the \atlas\ Combined Array for Research in Millimetre-wave Astronomy (CARMA) survey. For NGC0524 we use the PdBI observations of CO(1-0) from \cite{Crocker:2011ic}. For NGC3665 and NGC4526 we utilise the higher resolution CO(2-1) observations of {Onishi et al., in prep} and \cite{2013Natur.494..328D}, respectively. We refer readers to these paper for a full description of the data acquisition and reduction. In this analysis we utilise the full, cleaned CO data cubes created by the above authors. For NGC4459 we make use of new high resolution CARMA CO(2-1) data, and we describe these observations in Section \ref{newdata} below.

\begin{figure*}
\begin{minipage}{0.5\textwidth}
\includegraphics[height=8.25cm,angle=0,clip,trim=0cm 0cm 0cm 0.0cm]{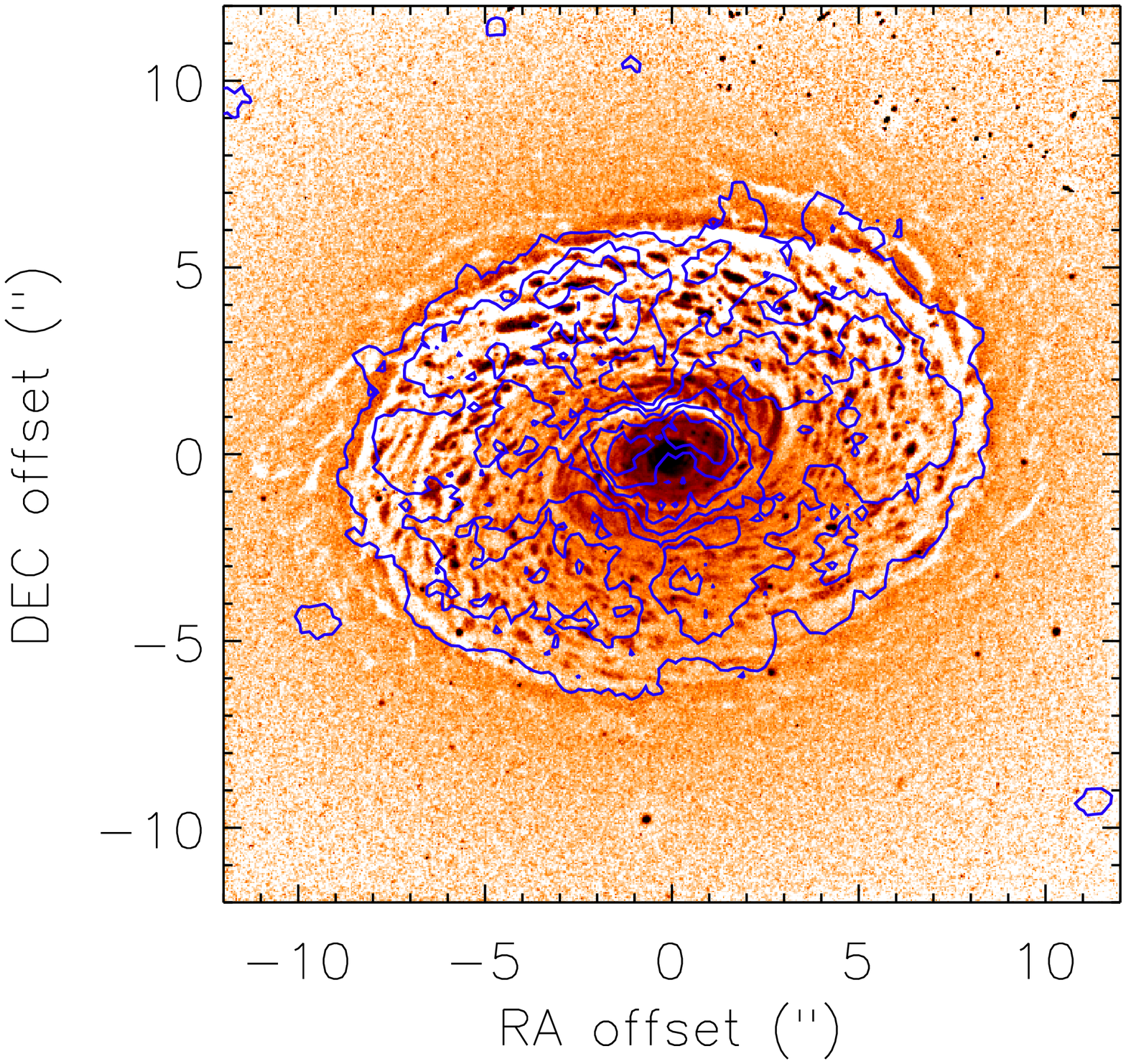}
\end{minipage}
\begin{minipage}{0.49\textwidth}
\includegraphics[height=3.5cm,angle=0,clip,trim=0cm 0cm 0cm 0.0cm]{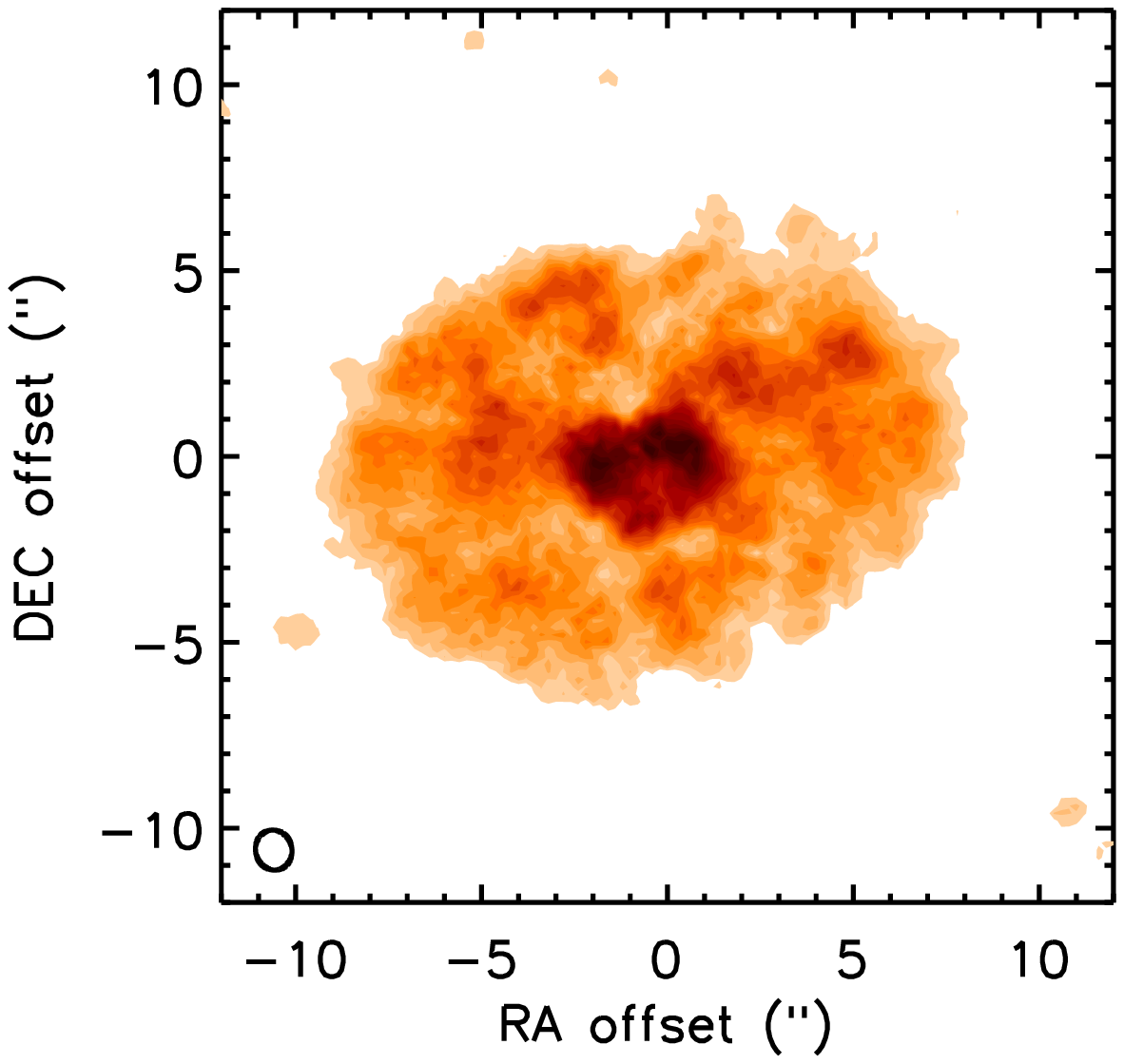}
\includegraphics[height=3.5cm,angle=0,clip,trim=2.25cm 0cm 0cm 0.0cm]{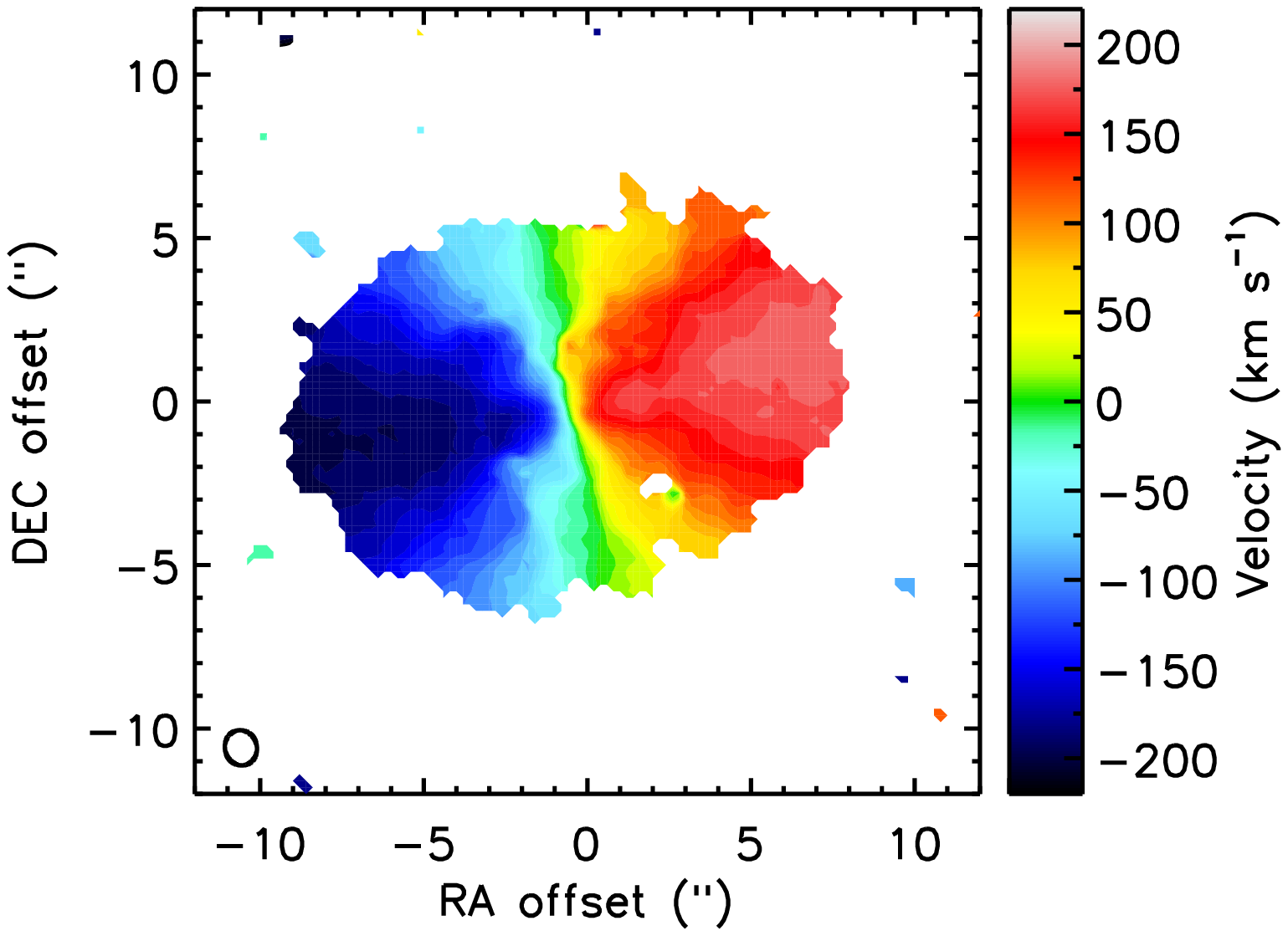}\\ 
\includegraphics[height=4.0cm,angle=0,clip,trim=0cm 0cm 0cm -0.5cm]{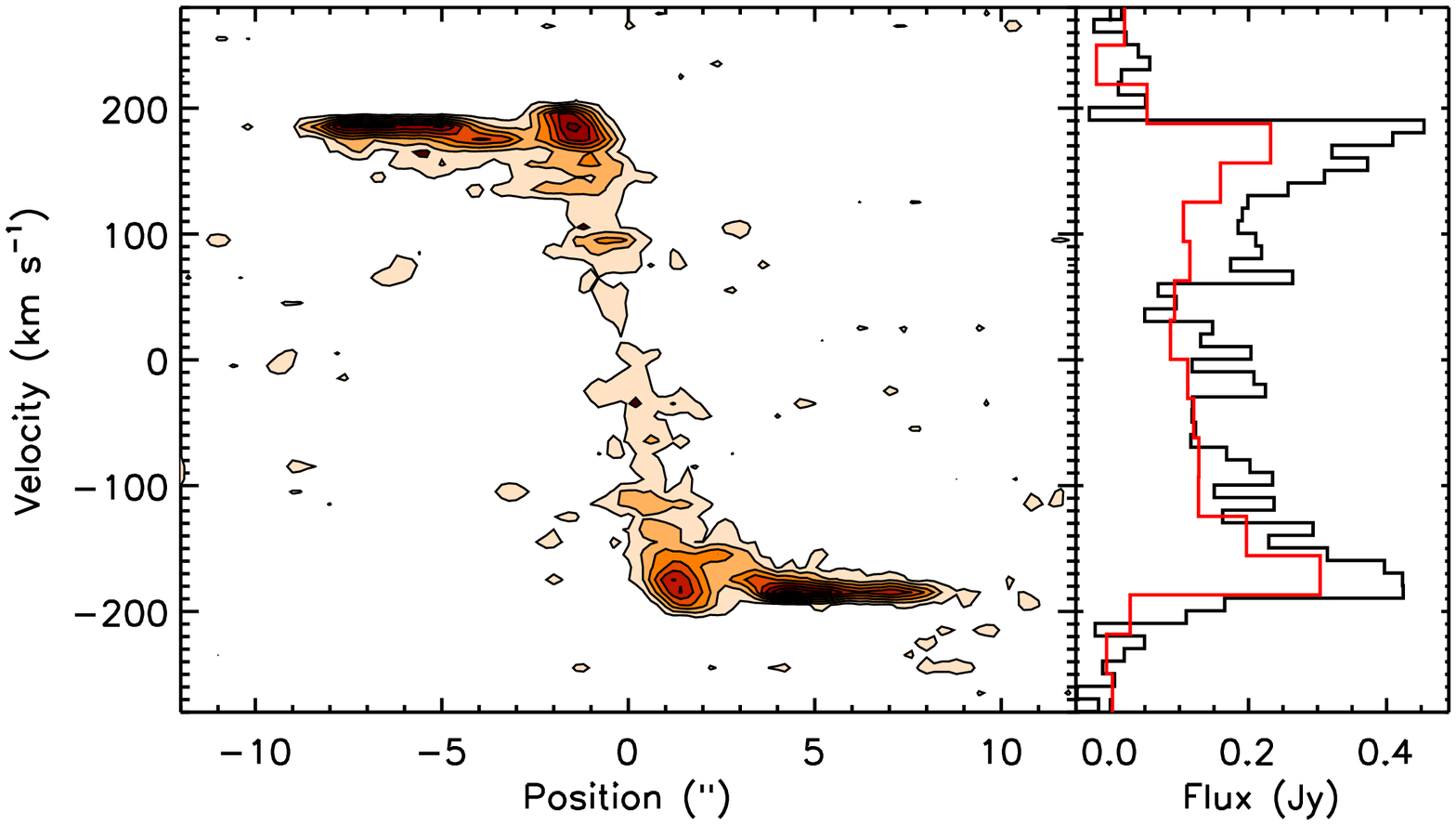}
\end{minipage}\vspace{0.5cm}
\includegraphics[width=1.0\textwidth,angle=0,clip,trim=0cm 0cm 0cm 0.0cm]{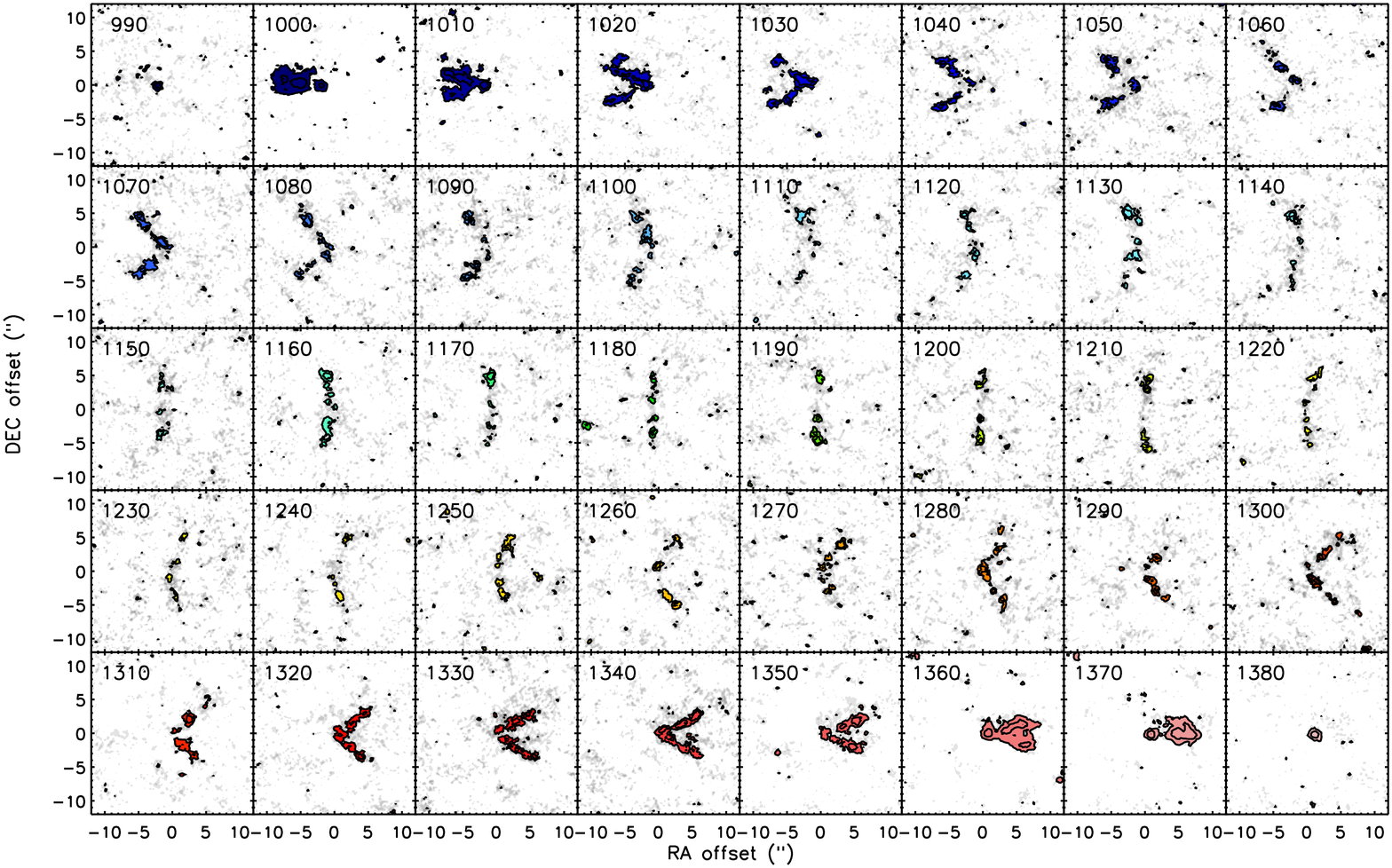}
\begin{center}
\caption{New CARMA observations of NGC4459. \textit{Top left:} Hubble Space Telescope UVIS F475W image of the galaxy centre, which has been unsharp-masked to show the dust disc. Overlaid are the contours of the CO(2-1) integrated intensity map.  \textit{Top right, upper panel:} Zeroth and First moment (Integrated intensity and velocity) maps of the CO(2-1) emission.  \textit{Top right, lower panel:} Position-velocity diagram of the CO emission extracted along the kinematic major axis. Also shown in the right panel is the total CO(2-1) spectrum of NGC4459 (black line). Overlaid on this in red is the IRAM-30m observation of \protect \cite{2007MNRAS.377.1795C}.  \textit{Bottom:} Greyscale shows the channel maps in the velocity interval where emission is detected (990-1380 \kms). The coloured region with black contours shows the areas detected with above a 3$\sigma$ significance.}
\label{4459data}
 \end{center}
 \end{figure*}

\section{CARMA observations of NGC4459}
\label{newdata}

In this study we utilise new high resolution CARMA data for the fast-rotating ETG NGC4459.
$^{12}$CO(2--1) observations of this object were taken in the B, C and D configurations of CARMA between the 25th November 2010 and the 3rd May 2011.  The combination of these arrays means we are sensitive to emission on scales of 0.3 -- 24\arcsec\ at the frequency of CO(2--1). Three 250 MHz correlator windows were placed over the CO(2--1) line, giving a continuous velocity coverage of $>$800\kms with a raw velocity resolution of $\approx$4.1 \kms. This is sufficient bandwidth and channel resolution to properly cover and sample the line. 

\subsection{Data reduction}
The raw CARMA visibility data were reduced using standard procedures, as documented in \cite{2013MNRAS.432.1796A}. We briefly summarise the salient details here. The Multichannel Image Reconstruction Image Analysis and Display ({\tt MIRIAD}) package \citep{1995ASPC...77..433S} was used the reduce the data.  For each track the raw data were Hanning-smoothed in velocity, and baselines that show decorrelation on calibrator sources were flagged.   
The data were then line-length corrected, and the bandpass was calibrated using observations of bright quasar 3C273.
 The atmospheric phase offsets present in the data were determined using 3C273 as a phase calibrator, observed at regular ($\approx15$ minutes) intervals.  Amplitude calibration was performed using Mars. Flux calibration uncertainties are assumed to be up-to 20\%. 
After the data were satisfactorily processed, the gain solutions derived from the nearby calibrator were applied to the source.  

We then used {\tt MIRIAD} to combine and image the resultant visibility files for each track, producing a three-dimensional (3D) data cube. The $uv$-datasets were transformed into RA-Dec-velocity space (with velocities determined with respect to the rest frequency of the CO(2-1) line). We here use data with a channel width of 10 \kms. Pixels of 0\farc2$\times$0\farc2 were chosen as a compromise between spatial sampling and resolution, typically giving approximately 5 pixels across the beam major axis.  One arcsecond corresponds to a physical scale of $\approx$78 pc in this source. We imaged the area within the primary beam of the 10m antennas ($\approx 54''$). The {\tt MIRIAD} imaging task {\tt INVERT} was run with the mosaicking option, to properly scale the data and account for the different primary beam widths.

The data presented here was produced using natural weighting, yielding a synthesised beam of 1\farc08$\times$0\farc98 (a linear resolution of $\approx$84$\times$76 pc). The dirty cubes were cleaned in regions of source emission to a threshold equal to the RMS of the dirty channels.  The clean components were then added back and re-convolved using a Gaussian beam of full-width at half maximum (FWHM) equal to that of the dirty beam.  This produced the final, reduced and fully calibrated data cube for NGC4459, which has an RMS (root-mean square) noise level of 4.21 mJy/beam. 

\subsection{Data products}

The clean fully calibrated data cube was used to create moment maps: a zeroth moment (or integrated intensity) map, and a first moment (mean velocity) map.
 In order to create these, the clean data cube was Hanning-smoothed in velocity and Gaussian-smoothed spatially (with a FWHM equal to that of the beam), and masks were created by selecting all pixels above a fixed flux threshold of 2.5$\sigma$, adjusted to recover as much flux as possible in the moment maps while minimising the noise.  The moment maps were then created using the unsmoothed cubes within the masked regions only.
 In addition, the major axis position velocity diagram (taken with a PA of 269$^{\circ}$, as determined in \citealt{2011MNRAS.417..882D}) and total spectrum were extracted.
These data products are shown in Figure \ref{4459data}.

\subsection{Gas morphology}

Figure \ref{4459data} shows the morphology of the gas in NGC4459. We detect gas in an oval region, coincident with the dust visible in HST imaging. The dust is distributed in a flocculent disc, with a sharp outer edge at a semi-major axis of 8\farc5 and only a few faint spiral dust features exist beyond this. The CO in this object also appears to be truncated at the outer edge of the dust disc, although the faint spiral features in the north west of the galaxy do have some detected emission. The inclination of 47$^{\circ}$ estimated by previous studies (e.g. \citealt{2008ApJ...676..317Y,2011MNRAS.414..968D}) would be fully consistent with the gas being distributed in a flat, circular disc in the galaxy plane.  Kinematic evidence also suggests that the gas in this object is distributed in a regularly rotating disc. The "spider diagram" seen in the channel maps is regular and symmetric, and (as our later analysis will show) can be used to predict an identical inclination to that estimated from photometry. 
 The molecular gas disc is flocculent, with various gas peaks that likely relate to marginally resolved giant molecular clouds within the disc. Unlike in lenticular galaxy NGC4526 \citep{2013Natur.494..328D,2015ApJ...803...16U}, however, some molecular gas emission is present at all positions in-between these peaks, suggesting the presence of a large population of unresolved clouds.
 
In the HST image one can see that from a radius of about 3\farc75, two dusty spiral arms connect the main disc with an inner dust ring of semi-major axis $\approx$2\farc6. \cite{2008ApJ...676..317Y} conclude that these features are trailing spirals. The arm features do not seem to be associated with strong peaks in the gas emission, however they do seem to affect the gas within the dust ring. Three peaks of emission are detected within the inner ring, which seem to form a gas ring when connected, with outer radius traced by the dust structure. Two of these peaks have approximately similar strength, and lie upstream of the point where the spiral dust structures connect to the ring. These peaks could be caused by buildup of material, as gas rotating around the ring meets material inflowing along the spiral structures. This hypothesis is supported by the line of sight CO spectra extracted from the position of these peaks, which have wings towards low velocity (i.e. towards the galaxy systemic) which cannot be reproduced by our kinematic modelling assuming circular orbits. The other peak to the south of the inner ring is less bright, and may be a single giant molecular cloud/association flowing around the ring.

\subsection{Comparison with existing observations}

\cite{2007MNRAS.377.1795C} observed CO(1-0) and CO(2-1) in NGC4459 with the IRAM-30m telescope, finding a total CO(2-1) flux of 86.5$\pm$3.1 Jy \kms. We compare the total spectrum extracted from our observations to the CO(2-1) data of these authors in Figure \ref{4459data}, finding that the total flux is $\approx$40\% higher in our interferometric data. This likely arises because the beam size of the IRAM-30m at 230GHz is 11\arcsec, similar to the total size of the molecular gas disc. Making mock observations of our resolved datacube using a gaussian beam shape of this size yields an almost identical spectrum that found by \cite{2007MNRAS.377.1795C}. We are hence confident that we have not resolved out significant flux in the central parts of this object. 

Resolved observations of the CO(1-0) line in this object were presented in \cite{2008ApJ...676..317Y}, with a beamsize of  9\farc0$\times$5\farc5. Our observations are entirely consistent with those of \cite{2008ApJ...676..317Y}. Convolving our observations with an asymmetric beam to match those in \cite{2008ApJ...676..317Y} yields very similar moment maps (and a matching major axis position-velocity diagram). We thus conclude that that CO(1-0) and CO(2-1) transitions arise from gas which is entirely co-spatial in this object, with identical kinematics. This is expected, given the low excitation temperatures of these transitions (which means they should trace the bulk of the molecular material in high metallicity objects like NGC4459).

\section{Method}
\label{method}
In this work we aim to constrain the mass profile of our target ETGs, and thus their mass-to-light ratio gradients. In order to do this we model the rotation of the molecular gas. We also require an estimate of the stellar population properties, and how these vary with radius. In this Section we explain the methods used to do this, dealing with molecular gas kinematic modelling in Section \ref{molmodel}, and stellar population analyses in Section \ref{stelpop}. Combining these two elements allows us to estimate the IMF mismatch parameter (Section \ref{imfmismatchsec}), and how it varies radially, and with respect to local galaxy properties (Section \ref{results}). 

\subsection{Molecular gas kinematic modelling}
\label{molmodel}
In order to estimate the mass-to-light ratio as a function of radius in our early-type sources we used a forward modelling approach. We utilised the KINematic Molecular Simulation (KinMS\footnote{available at https://github.com/TimothyADavis/KinMS}) mm-wave observation simulation tool of \cite{2013MNRAS.429..534D}. 
This tool allows us to input guesses for the true gas distribution and kinematics, and (taking into account the observational effects of beam-smearing, disc thickness, velocity dispersion, binning, etc) produce a simulated datacube which can be compared with the observed data. 

In order to determine the rotation speed of the gas as a function of radius, as in \cite{2014MNRAS.443..911D} we used the Markov Chain Monte Carlo (MCMC) code \textsc{KinMS\_mcmc} that couples to the KinMS routines, and allows us to fit the data and obtain the full bayesian posterior probability distribution for the fitted parameters. This code fits the entire data cube produced by the interferometer, rather than simply the position-velocity diagram (as was done in \citealt{2013Natur.494..328D}). 
 The simulations used a beam, pixel size and velocity resolution identical to our observations.

\subsubsection{Gas distribution and kinematics}
For each object we created a simple model of the gas distribution, which is then used as an input to the modelling code. We do not aim to reproduce all the internal features of gas distribution, only the broader details of at which locations gas is present or absent. 
 In NGC0524 and NGC3607 we chose to distribute the gas in an exponential disc. This simple, zeroth order form (which has been shown to be appropriate in most ETGs; \citealt{2013MNRAS.429..534D})  provides a good match to the observed morphology of the gas in these objects.
In NGC3665 and NGC4429 we used exponential discs with an inner cutoff (as explained in {Onishi et al., in prep}). In NGC4459 we again used an exponential disc model with an inner cutoff, with an additional gaussian enhancement of the gas density at the radius of the molecular ring seen in our observations (see Section \ref{newdata}). In NGC4525 we use a three ring model, as previously presented in \cite{2014MNRAS.443..911D}.  We note, however, that (as explored in detail in \citealt{2014MNRAS.443..911D}) using a more simple form for the gas distribution does not bias our results. The exponential disc and ring radii (and cutoff radii) were all left as free parameters, which were modified by the \textsc{KinMS\_mcmc} code to obtain a good fit.

In addition to the free parameters related to the gas distribution, various other free parameters of the gas disc are included in the model. These are the total flux, position angle, and inclination of the gas disc, as well as its kinematic centre (in RA and Dec) and its systemic velocity. We find no evidence of disc warps in the sample galaxies, so the inclination and position angle are fitted as a single value, which applies in each radial bin.

 We assume implicitly that the gas in in circular rotation within the potential, and hence the gas velocity varies only radially. We parameterise the kinematics in radial annuli, of one beamwidth across. As our sources have various physical extents and our observations have different beamsizes, we are able to fit between 3 and 10 independent radial bins per source. In addition to the free parameters controlling the gas rotation curve, we also include a free parameter for the internal velocity dispersion of the gas disc, which is assumed to be constant radially within our objects. For those galaxies with high resolution data (NGC0524, NGC3665, NGC4526) we include the gravitational effect of the measured SMBH in our model, with its mass fixed to the value found by other authors (\citealt{2009MNRAS.399.1839K}, Onishi et al., in prep , \citealt{2013Natur.494..328D}). We then allow the MCMC code to estimate the circular velocities within each of our bins (see Section \ref{fitting} for more details).

In order to convert the derived velocity profiles to dynamical mass-to-light ratio measurements we here parameterise the luminous matter distribution using multi-Gaussian expansion (MGE; \citealt{Emsellem:1994p723}) models of the stellar light distribution. These were constructed from HST images (at the longest wavelength available, in order to minimise dust contamination), and from $r$-band SDSS images in cases where this was not possible. In IC719 we use Spitzer space telescope 3.6$\mu$m observations, as the source is very dusty.
Assuming the gaussian density distribution assumption made when constructing an MGE holds, this model of the stellar light can easily be de-projected using the observed inclination (as fitted by \textsc{KinMS\_mcmc}). It then directly predicts the circular velocity of the gas caused by the luminous matter, modulo the stellar mass-to-light ratio. 

At each step of the MCMC chain we use the MGE to predict the rotation curve of the galaxy at some inclination, imposing an M/L of unity. Then using the basic relationship

   \begin{eqnarray}
V=\sqrt{\frac{GM}{r}}=\sqrt{\frac{G\left(L\frac{M}{L}\right)}{r}}=\sqrt{\frac{GL}{r}}\sqrt{\frac{M}{L}},\label{massinceq}
\label{vpropml}
\end{eqnarray}
we can multiply the derived rotation curve by the M/L factor chosen at this step for each radial bin. We define this value at the bin centre radius, and interpolate linearly between bins. Using stepwise transitions, or higher order splines to interpolate between points does not change our results. Note that although Equation \ref{massinceq} is only formally valid for a spherical mass distribution, the proportionality is valid generally.

In this way, comparing the derived rotation curve with that expected from the luminous matter alone allows us to estimate the dynamical mass-to-light ratio at each radius, which can then be compared with that predicted from stellar population analyses.

\subsubsection{Fitting process}
\label{fitting}
In order to ensure our kinematic fitting process converges we set reasonable priors on some of the parameters. The kinematic centre of the galaxy was constrained to lie within one beam-width of the optical galaxy centre position (but good fits are always found well within this). The systemic velocity was allowed to vary by $\pm$50 \kms\ from that found by optical analyses (but again good fits are always found in the inner part of this range). The gas velocity dispersion was constrained to be less than 50 \kms (but is always found to be low, $<$12 \kms), and the disc scale length/cutoffs constrained to be less than 20\arcsec. The M/L prior in each radial bin allows variation between 0.1 and 20. The inclination of the gas disc is allowed to vary over the full physical range allowed by the MGE model.
 A flat prior was used on each of these parameters (an assumption of maximal ignorance). 

Once the MCMC chains converged we ran the final iteration 30,000 times (with a 10\% burn-in) to produce our final posterior probability distribution.  These probability surfaces were then marginalised over the other parameters in order to produce an estimate of each parameter, and its associated 1 and 3$\sigma$ errors.

\subsection{Stellar population analysis}
\label{stelpop}
Using the publicly available spectral data cubes from the Atlas3D Survey\footnote{Available from www.purl.org/atlas3d}, we derive the spatially-resolved stellar mass-to-light ratio from the integrated stellar light using spectral fitting. Spectra are first binned to a minimum signal-to-noise ratio of 80, using the Voronoi tessellation method of \cite{2003MNRAS.342..345C}. The penalised pixel fitting (pPXF) algorithm of \cite{2004PASP..116..138C} was then used, together with simple stellar population (SSP) templates taken from the MIUSCAT model library \citep{2012MNRAS.424..157V}. Regions of the spectra potentially affected by emission lines (namely H$\beta$ and [OIII]) are excluded from the fit. The selected templates span 0.1-14 Gyr in age, and -1.71-0.2 in [Z/H], giving 264 simultaneously fitted templates. We employ linear regularisation constraints as per \cite{2015MNRAS.448.3484M} to obtain smooth weight distributions in the two dimensional space of age and metallicity. The resulting weights, $w_i$, are used to compute the mass-weighted M/L as:

\begin{equation}
M/L_{\rm pop,salp} = \frac{\sum w_i M_{{\rm *+rem},i}}{\sum w_i L_{X,i}}
\end{equation}

\noindent where $M_{{\rm *+rem},i}$ is the mass existing in stars and remnants, and $L_{X,i}$ is the luminosity in the given filter passband $X$, both corresponding to the age and metallicity associated with weight $w_i$, and assuming a unimodal power law IMF of the form: $\zeta(m) \propto m^{-2.35}$ \citep{1955ApJ...121..161S}. This procedure is fully consistent with that employed in \cite{2012Natur.484..485C}.

{M/Ls for photometric bands falling outside the MIUSCAT spectral range come from the photometric predictions of \cite{1996ApJS..106..307V}, available from the MILES website for ages and metallicities consistent with the MIUSCAT models. For overlapping  optical bands, these M/L predictions agree with the baseline MIUSCAT models with a standard deviation of 0.04. For the Spitzer 3.6$\mu$m band we use the MIUSCAT-IR models of \cite{2015MNRAS.449.2853R}. Note that these models only extend down to a metallicity ([Z/H]) of -0.4. For objects with significant low metallicity populations this means we may somewhat underestimate 3.6$\mu$m stellar population M/Ls \citep[see e.g.][]{2014ApJ...788..144M}.}

{Random errors on the individual M/L measurements were derived using Monte-Carlo simulations of the pPXF spectral fit {\it without} regularisation applied, and taking the standard deviation of the resulting M/L distribution. Typical M/L error values derived in this way are around $\pm$0.14 M$_{\odot}$/L$_{\odot}$.}

{To explore correlations of the spatially-resolved IMF and stellar population parameters, as has been done in the recent literature \citep[e.g.][]{2012ApJ...760...71C,2014ApJ...792L..37M,2015ApJ...806L..31M}, we employ line index measurements to derive the SSP-equivalent parameters of age, metallicity, and alpha-enhancement. We use a combination of Hbeta, Fe5015 and Mgb Lick index measurements cleaned for ionized gas emission lines using the above pPXF spectral fits, and find the best fitting single stellar population (SSP) model from \cite{2007ApJS..171..146S} that simultaneously reproduces these indices. A full comparison of how these SSP parameters compare with the mass-weighted parameters derived from spectral fitting is presented for these galaxies (as part of the larger \atlas\ Survey) in \cite{2015MNRAS.448.3484M}.}

{We note that these SSP-equivalent parameters are only used for exploring correlations with our inferred IMF mismatch parameter described below. The M/L$_{\rm pop,salp}$ itself is derived using spectral fitting, which allows us to account for the influence of a spread of ages and metallicities on the integrated light through the combination of fitted templates. Regularisation ensures a maximally smooth star formation history that still fits the data within the observational uncertainties, though we find that the resulting M/L is not significantly dependent on the degree of regularisation used - the average absolute difference between  the regularized and non-regularized M/L is 0.16, which is comparable to the random uncertainties.}

\begin{figure*}
\begin{center}
 \includegraphics[height=6cm,angle=0,clip,trim=0cm 0cm 0cm 0.0cm]{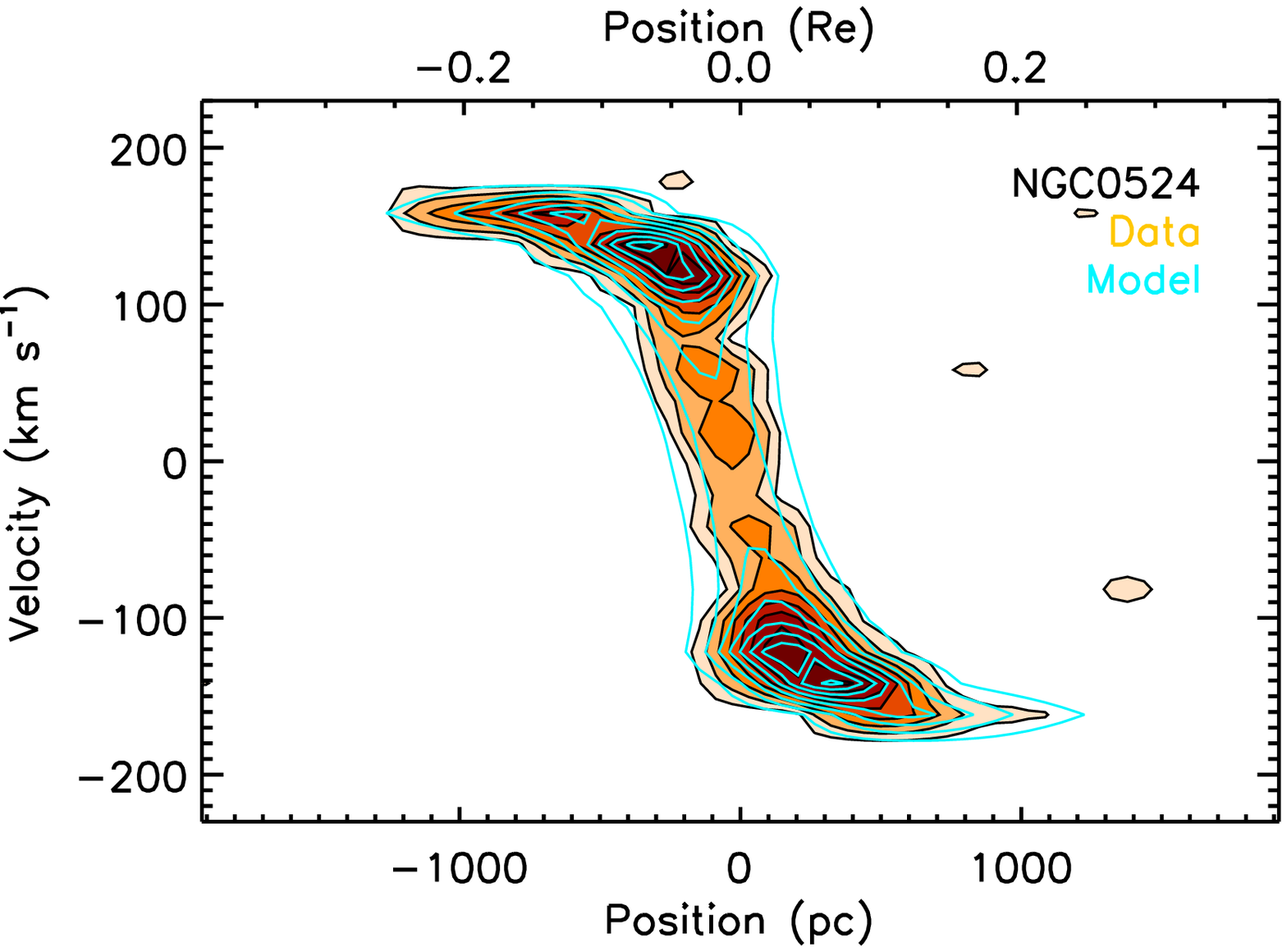}\hspace{0.5cm}
\includegraphics[height=6cm,angle=0,clip,trim=0cm 0cm 0cm 0.0cm]{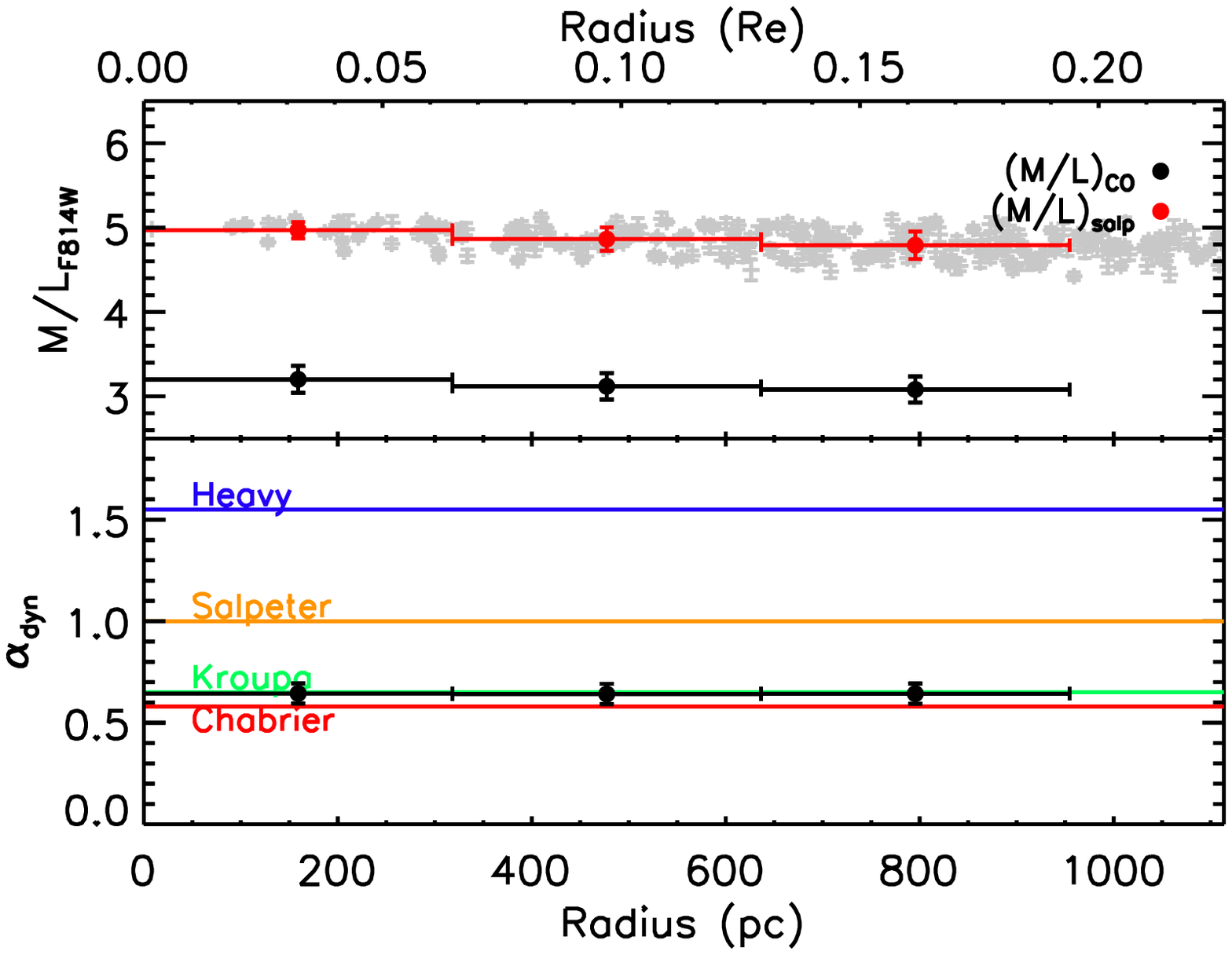}\\ \vspace{0.5cm}
\includegraphics[height=6cm,angle=0,clip,trim=0cm 0cm 0cm 0.0cm]{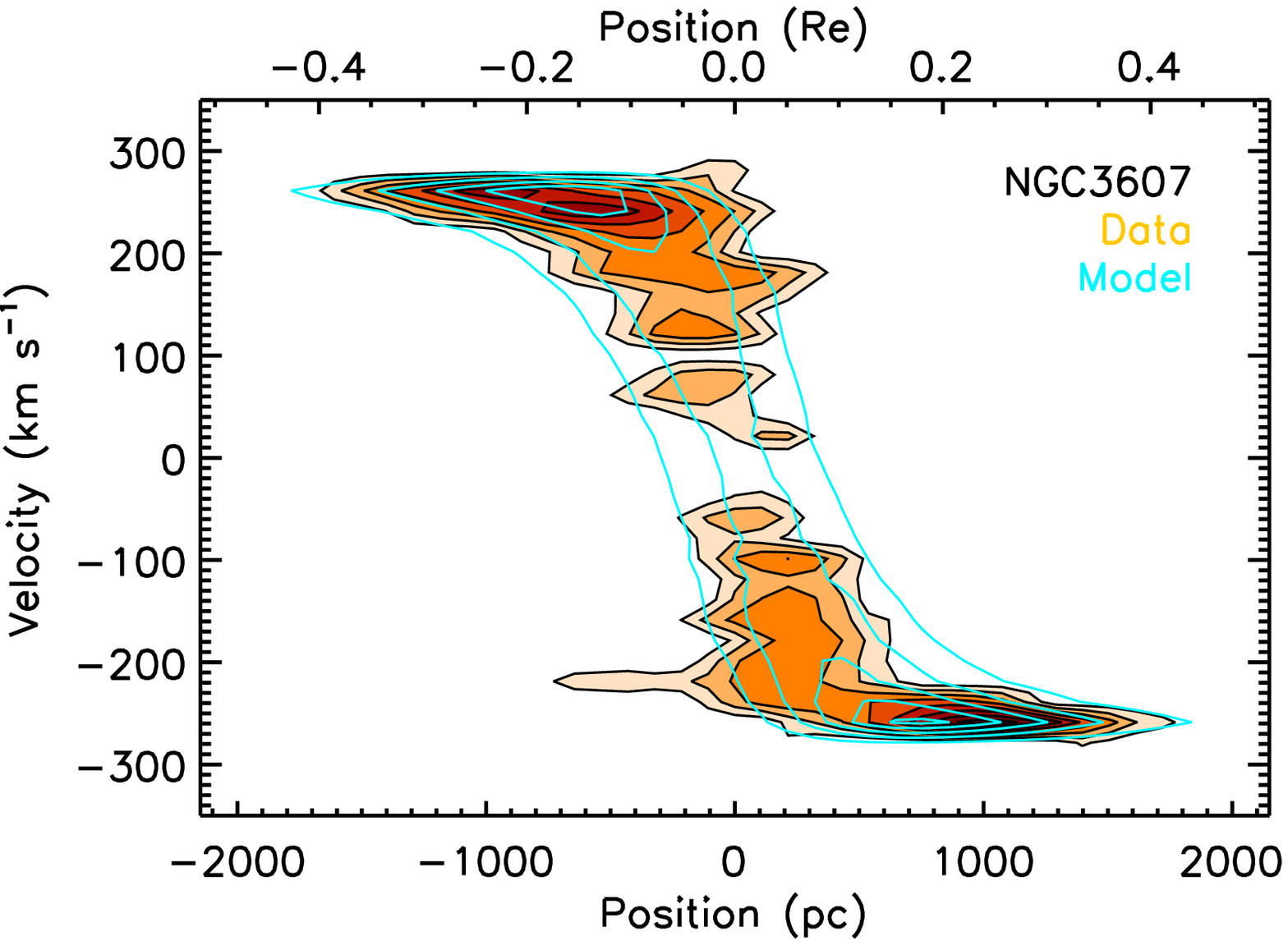}\hspace{0.5cm}
\includegraphics[height=6cm,angle=0,clip,trim=0cm 0cm 0cm 0.0cm]{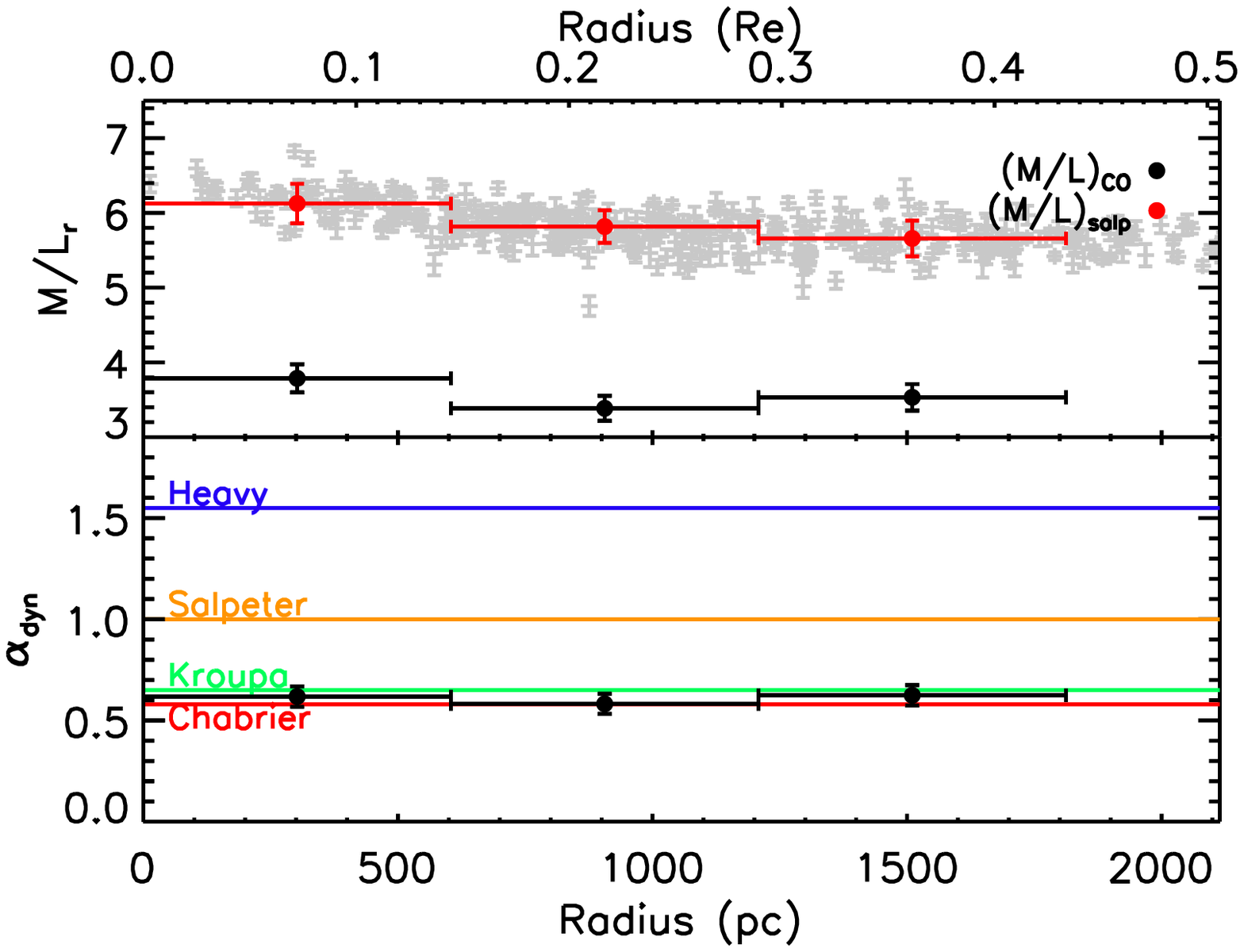}\\ \vspace{0.5cm}
\includegraphics[height=6cm,angle=0,clip,trim=0cm 0cm 0cm 0.0cm]{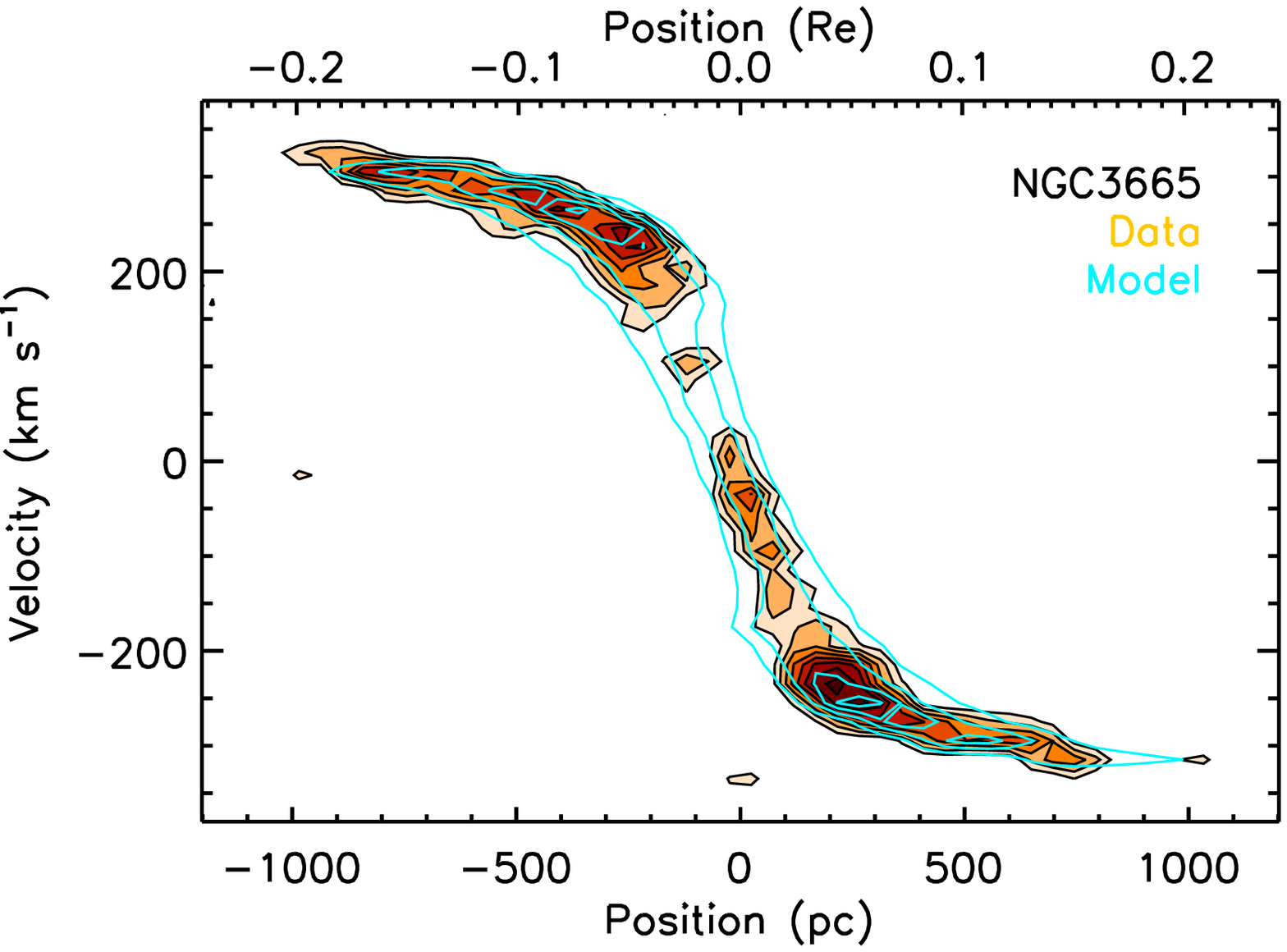}\hspace{0.5cm}
\includegraphics[height=6cm,angle=0,clip,trim=0cm 0cm 0cm 0.0cm]{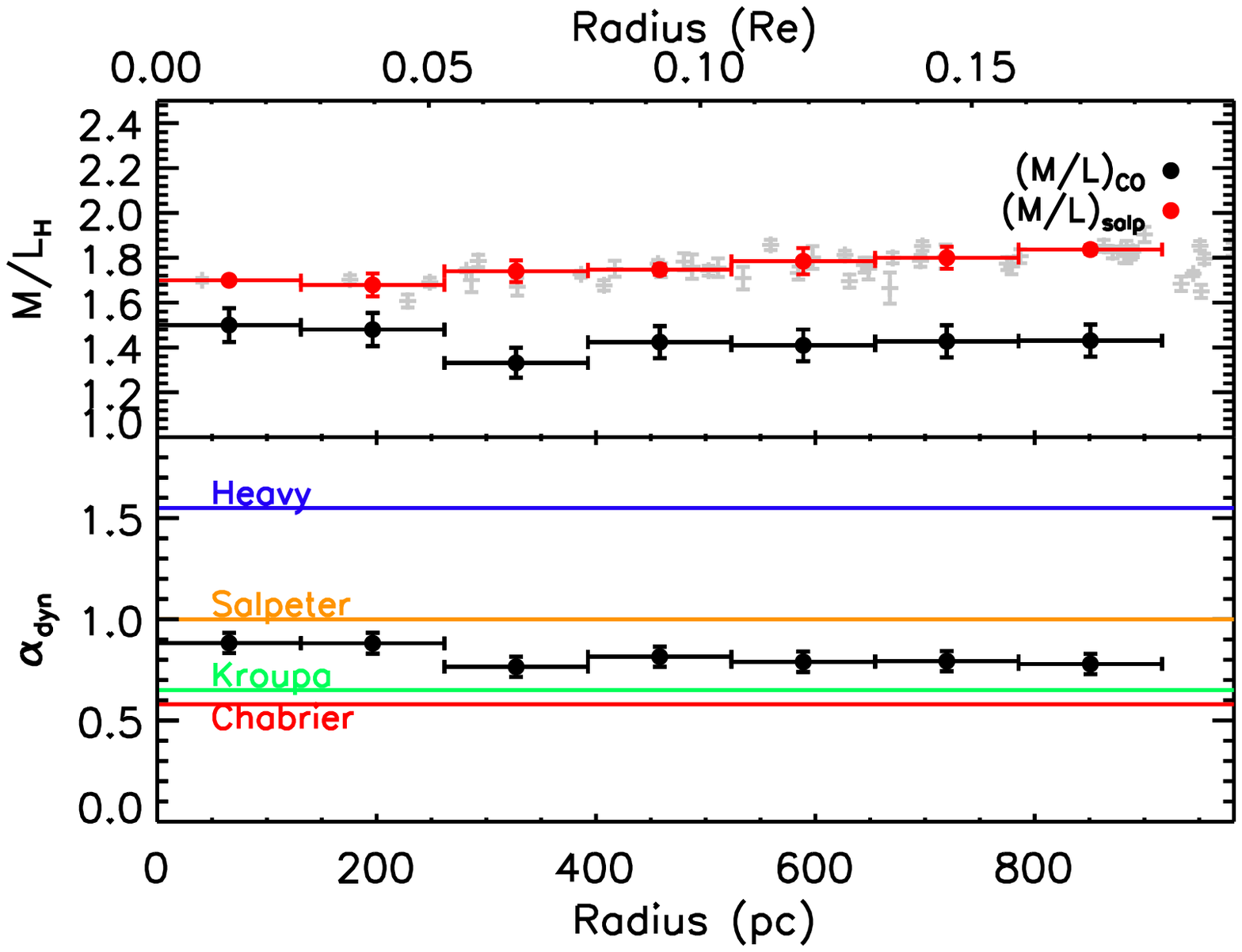}
 \caption{
 \textit{Left:} Observed major-axis position velocity diagram for each galaxy in our sample (black contours filled in orange), over-plotted in blue with the best fit position-velocity diagram extracted from our model (which was actually fitted on the full interferometric data cube) in an identical way. \textit{Right:}  The top panel shows the mass-to-light ratio gradient present within the source, as derived from the molecular gas observations in black. Also shown as grey plus symbols are the stellar population M/L values derived from the independent IFU bins at each radius. The mean value and scatter around this within each bin are shown as the red points with error bars.  In the bottom panel we show the IMF mismatch parameter in each bin (as defined in Equation \ref{imfmismatch}). For reference we show lines which correspond to Chabrier, Kroupa, Salpeter and Heavy IMFs.}
 \label{imffig}
 \end{center}
 \end{figure*}
    \begin{figure*}
\begin{center}
\includegraphics[height=6cm,angle=0,clip,trim=0cm 0cm 0cm 0.0cm]{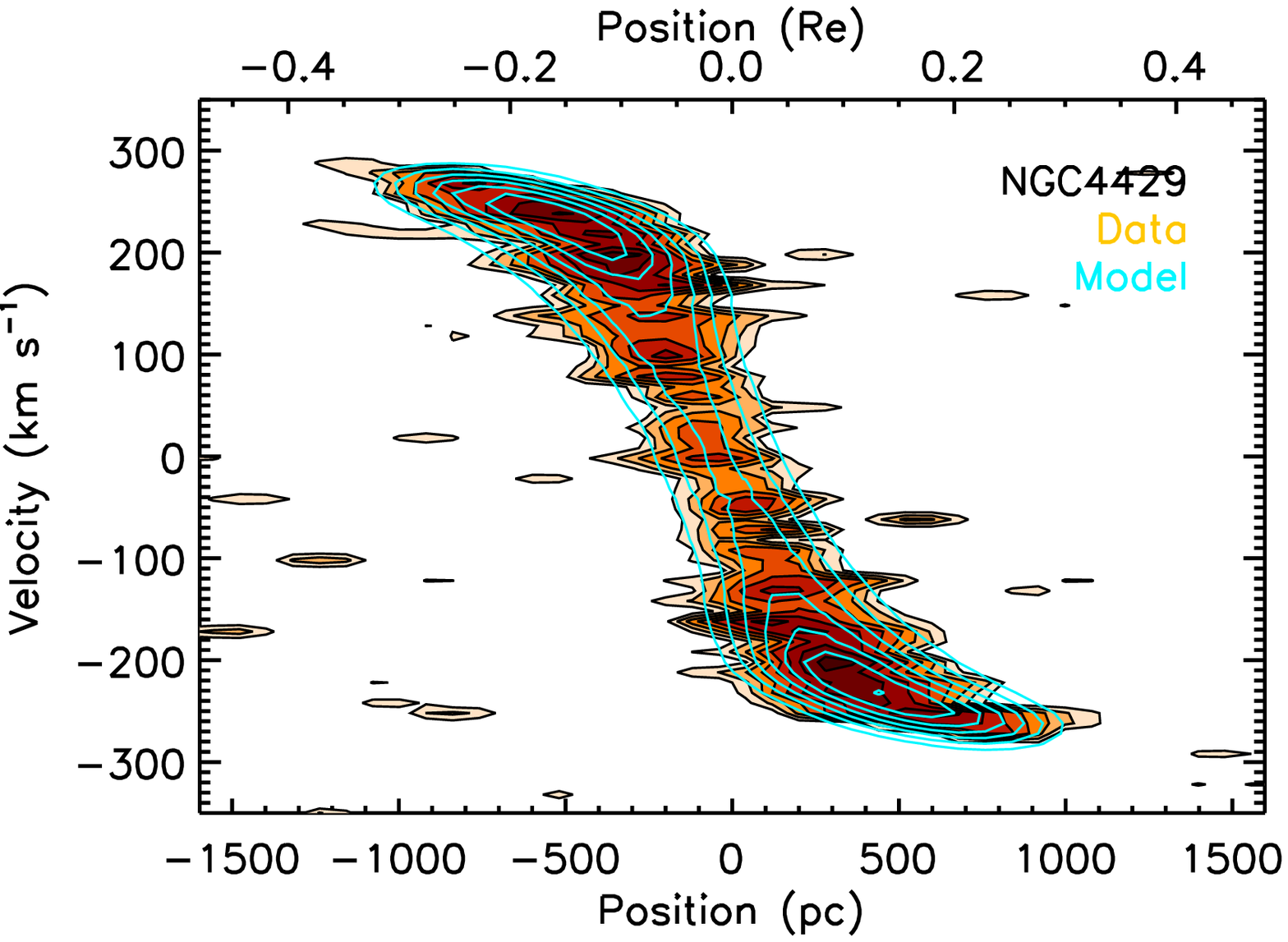}\hspace{0.5cm}
\includegraphics[height=6cm,angle=0,clip,trim=0cm 0cm 0cm 0.0cm]{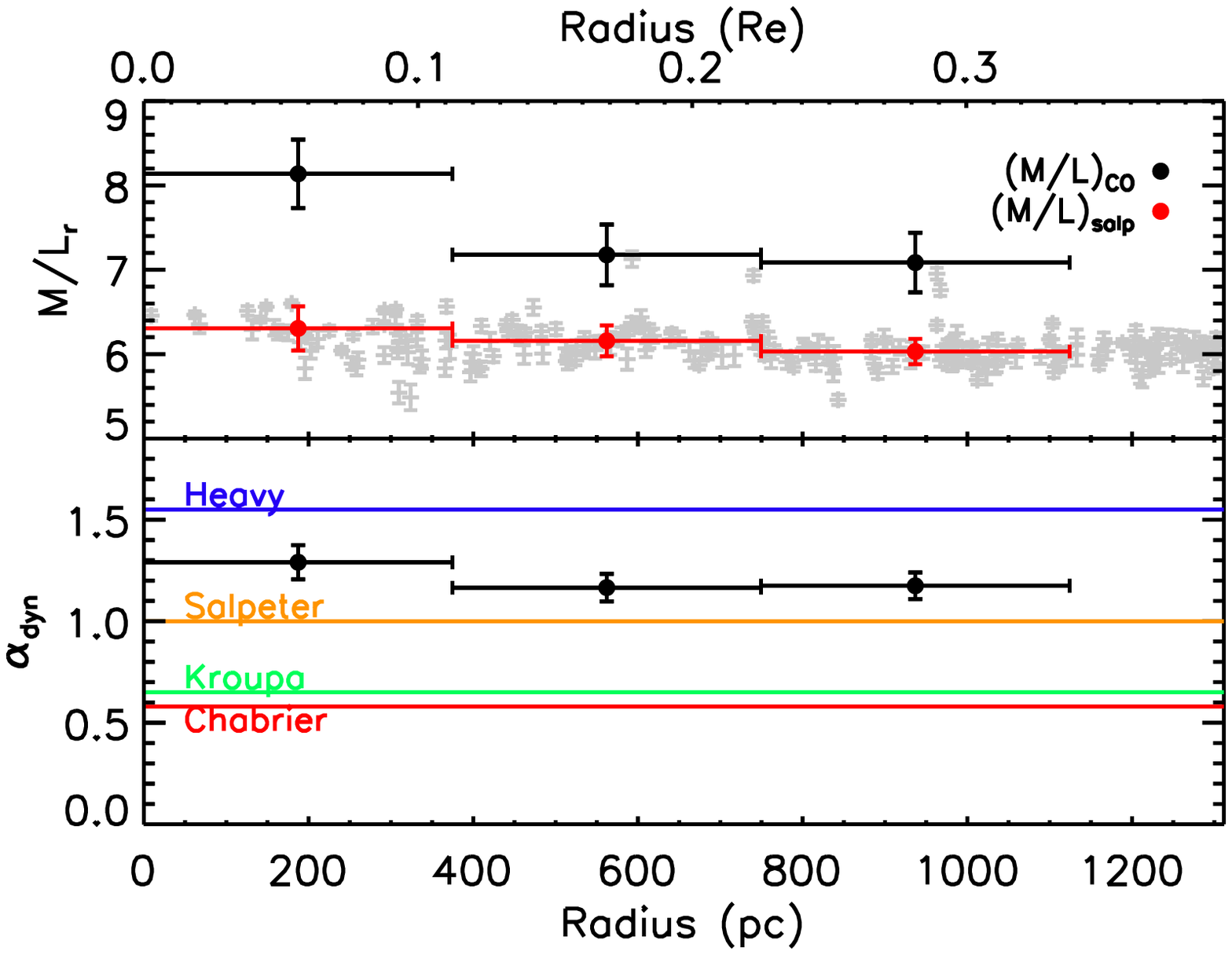}\\ \vspace{0.5cm}
\includegraphics[height=6cm,angle=0,clip,trim=0cm 0cm 0cm 0.0cm]{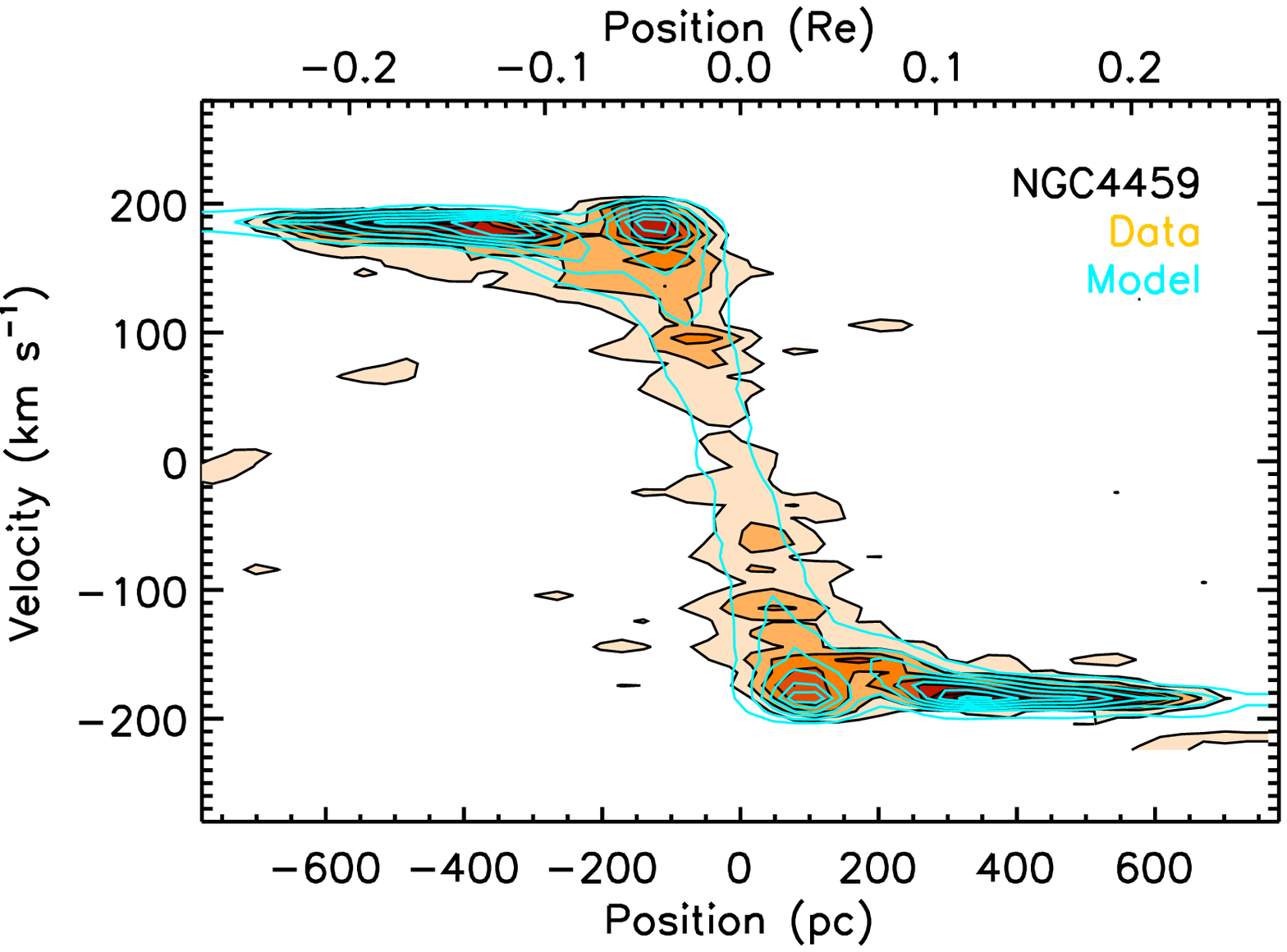}\hspace{0.5cm}
\includegraphics[height=6cm,angle=0,clip,trim=0cm 0cm 0cm 0.0cm]{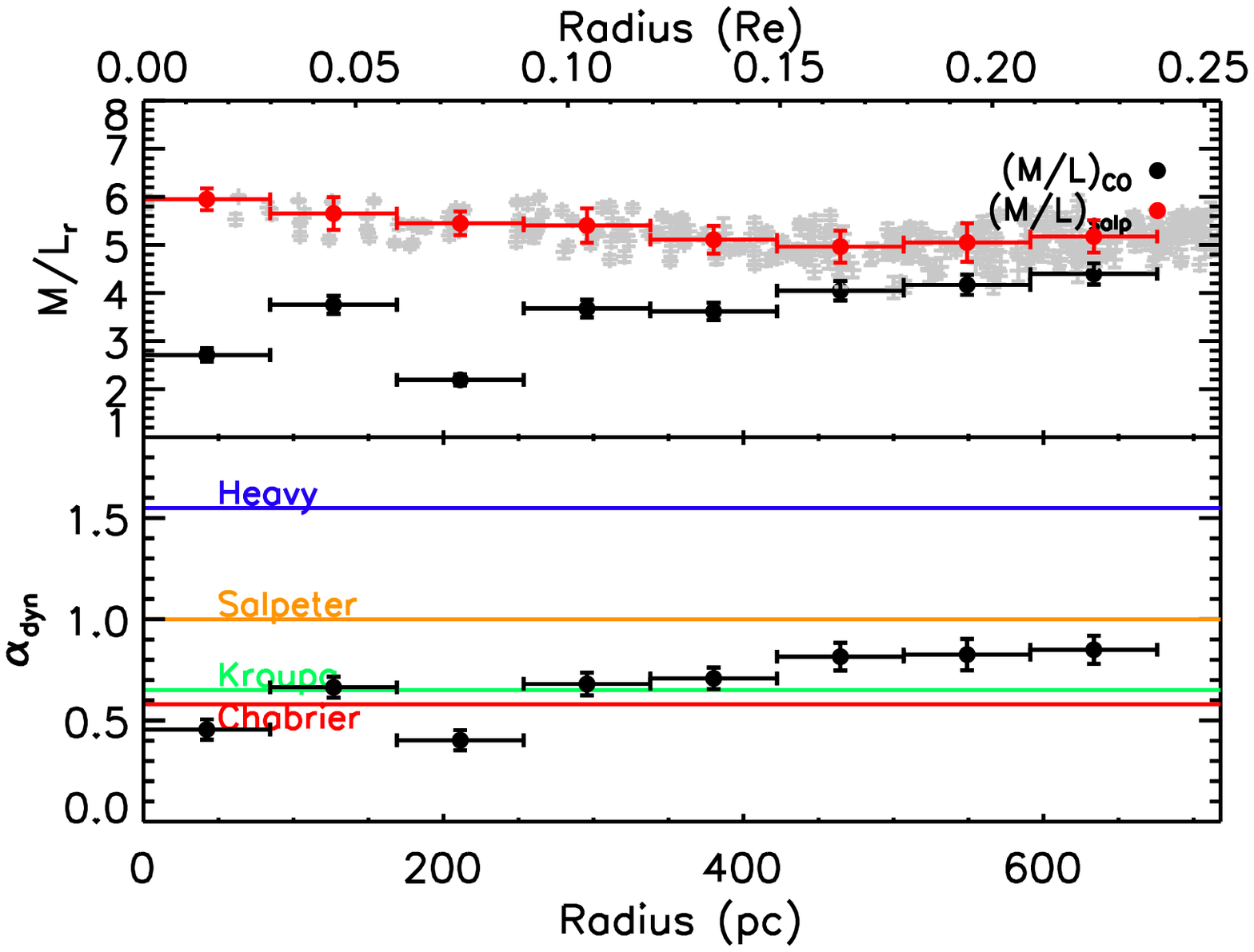}\\ \vspace{0.5cm}
\hspace{0.5cm} \includegraphics[height=6cm,angle=0,clip,trim=0cm 0cm 0cm 0.0cm]{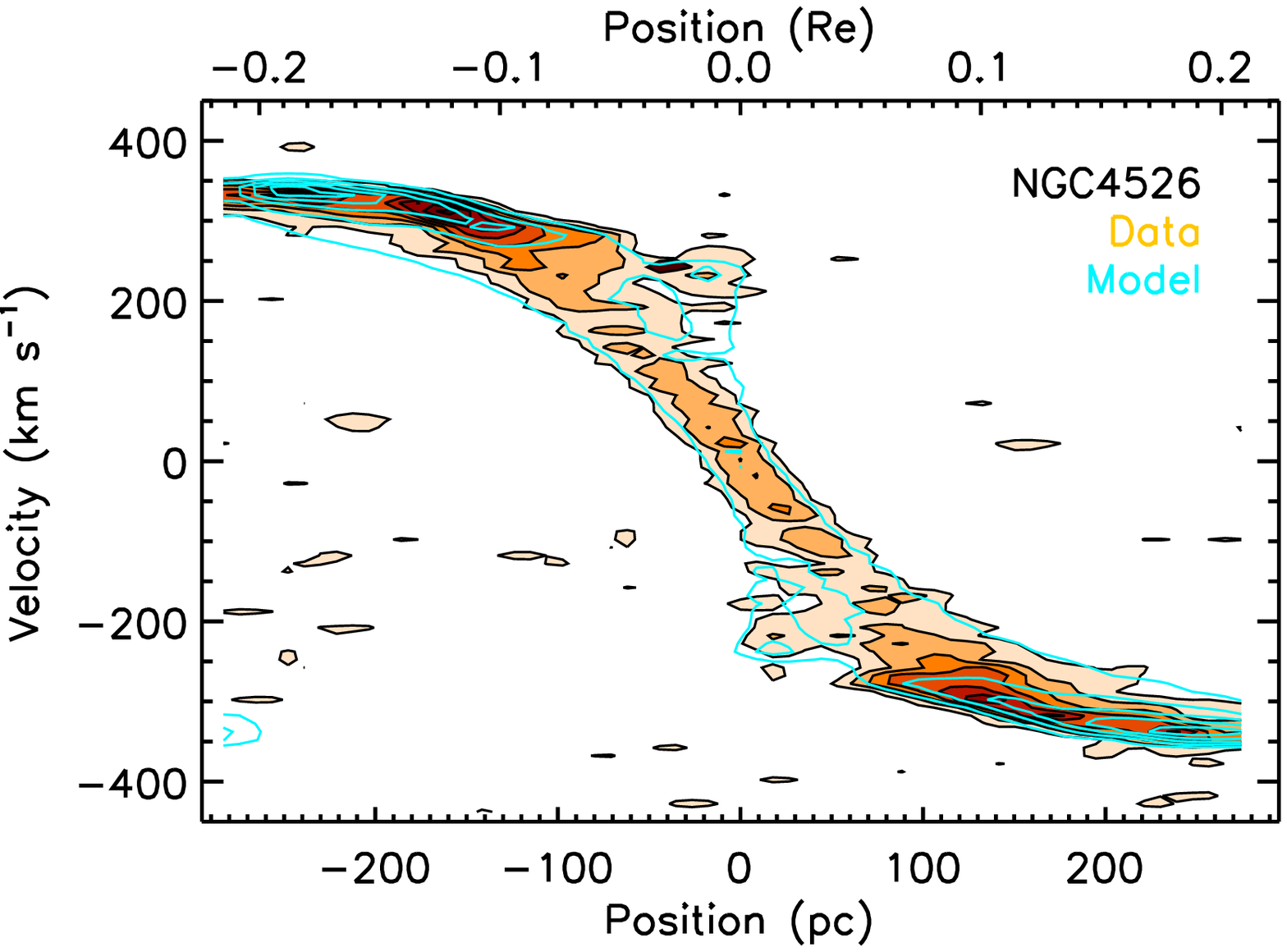}\hspace{0.5cm}
\includegraphics[height=6cm,angle=0,clip,trim=0cm 0cm 0cm 0.0cm]{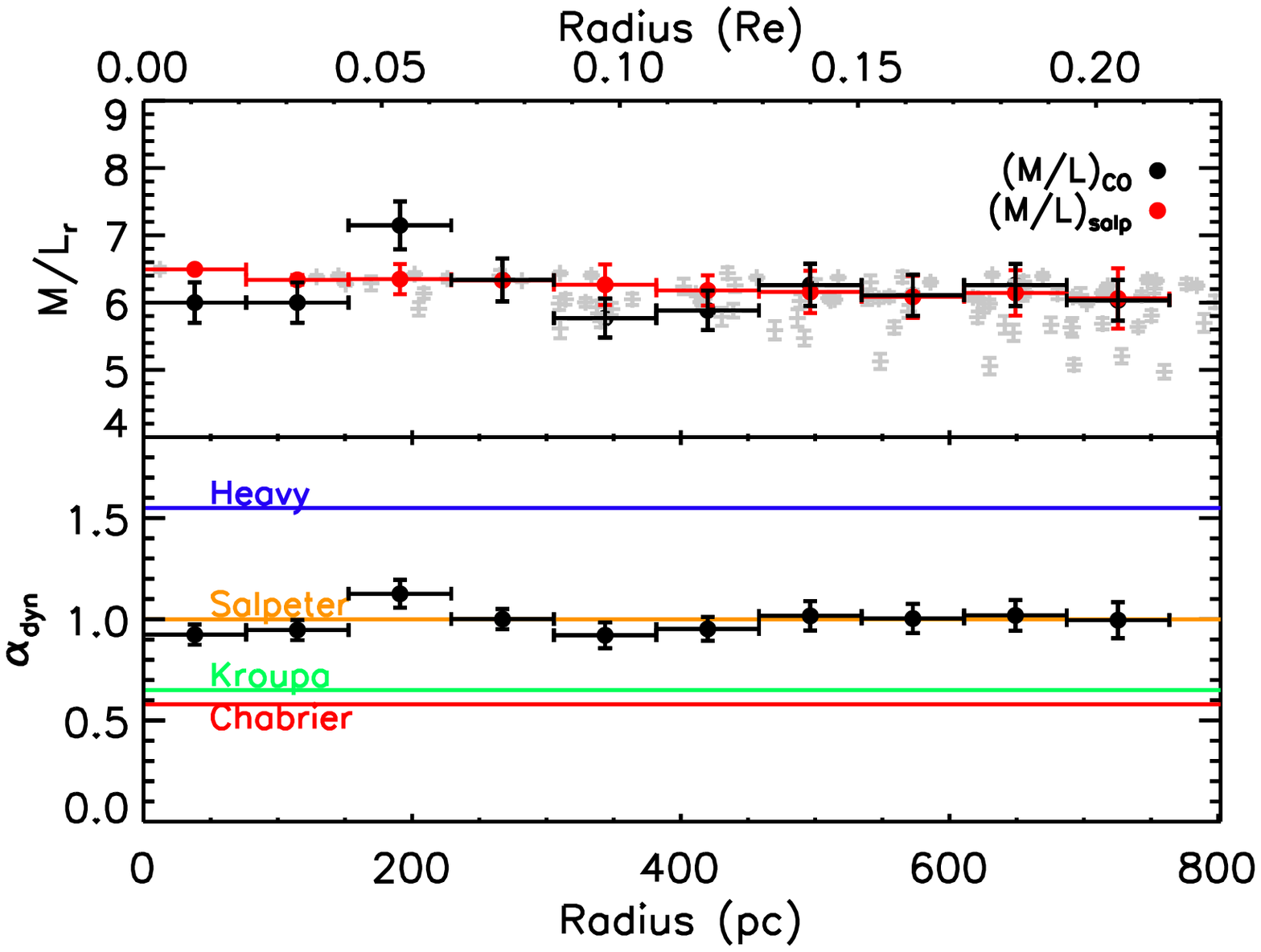}
 \contcaption{}{}
 \end{center}
 \end{figure*}
    \begin{figure*}
\begin{center}
\includegraphics[height=6cm,angle=0,clip,trim=0cm 0cm 0cm 0.0cm]{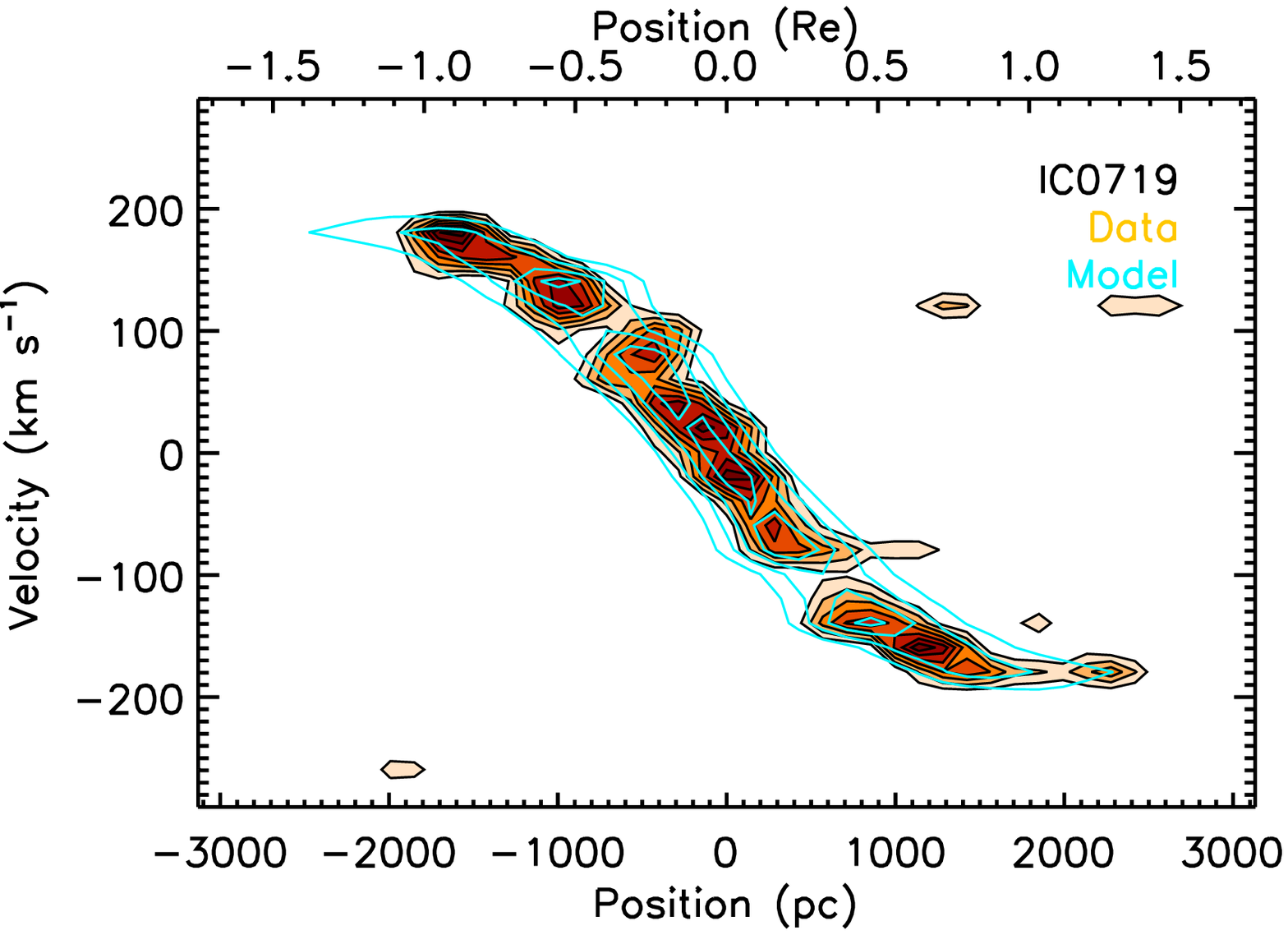}\hspace{0.5cm}
\includegraphics[height=6cm,angle=0,clip,trim=0cm 0cm 0cm 0.0cm]{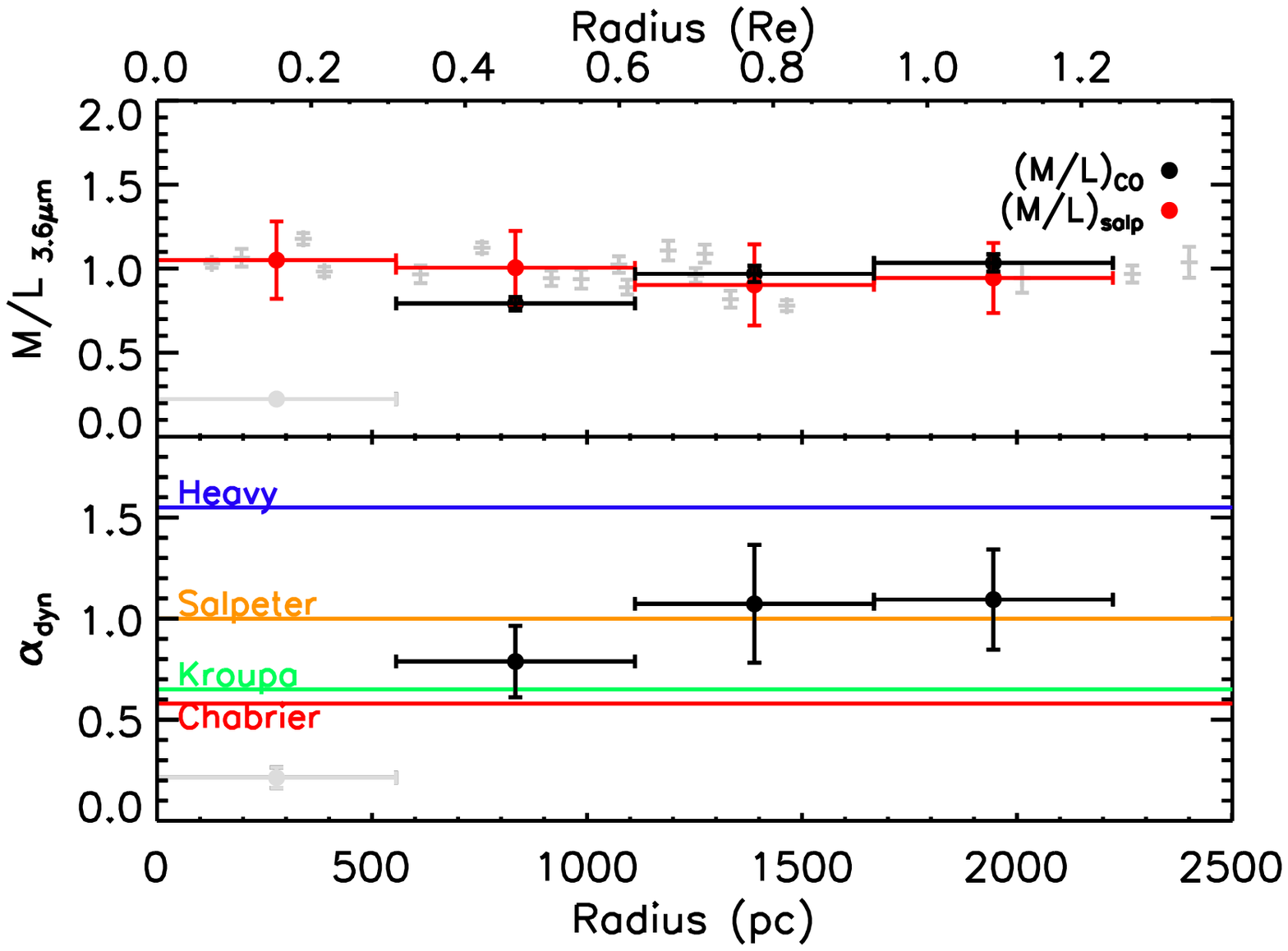} 
 \contcaption{}{}
 \end{center}
 \end{figure*}

\subsection{Derivation of IMF mismatch parameters}
\label{imfmismatchsec}
Following \cite{2012Natur.484..485C} we define the IMF mismatch parameter ($\alpha_{\rm dyn}$) as

\begin{equation}
\label{imfmismatch}
\alpha_{\rm dyn} = \frac{M_{\rm dyn}}{M_{\rm pop,salp}} = \frac{M/L_{\rm dyn}}{M/L_{\rm pop,salp}},
\end{equation}

\noindent where $M_{\rm dyn}$ is the dynamical mass within a bin estimated from the CO kinematics (which makes no assumption on the IMF), and $M/L_{\rm pop,salp}$ is the stellar mass-to-light ratio, estimated from the stellar population analysis assuming a \cite{1955ApJ...121..161S} IMF as described in the previous subsection. This ratio is equal to unity if the real IMF in the galaxy follows a \cite{1955ApJ...121..161S} form, 0.58 for a \cite{2003PASP..115..763C} IMF, 0.65 for a \cite{2001MNRAS.322..231K} IMF, and 1.55 for a Heavy IMF (as discussed in e.g. \citealt{2012Natur.484..485C}).

\section{Results}
\label{results}

In this section we present the best fitting M/L and $\alpha_{\rm dyn}$ gradients for our seven sample objects. We compare the values derived with those found by other authors, who used different techniques. We also explore some to the uncertainties that can affect measurements of this type in Section \ref{uncertainties}.

\subsection{M/L and IMF gradients}
For each object in our sample we present two plots to show our derived M/L and IMF gradients.
The first displays the observed major-axis position velocity diagram, over-plotted with the position-velocity diagram extracted from our model in an identical way. Although the model was actually fitted on the full interferometric data cube, this figure allows the reader to easily assess by eye the quality of the fits.

The second plot contains two panels. The top panel shows the mass-to-light ratio gradient present within the source, as derived from the molecular gas observations in black. Also shown as grey plus symbols are the stellar population M/L values derived from the independent IFU bins at each radius. The mean value and scatter around this within each bin are shown as the red points with error bars.  In the bottom panel we show the IMF mismatch parameter in each bin (the ratio of the black and the red point from the upper plot). For reference we show lines which correspond to Chabrier, Kroupa, Salpeter and Heavy IMFs. We discuss the fits obtained for each object individually below, and note any specific uncertainties individual to the source.

\subsubsection{NGC0524}

NGC0524 is a high-mass isolated fast-rotating ETG, orientated nearly face-on to our line of sight. It has a disc of molecular gas and dust that is clearly visible in optical imaging, but very little star formation. The molecular disc is regular and relaxed at our resolution, and its surface brightness profile is well fitted with an exponential disc model. We use the mass model from \cite{2009MNRAS.399.1839K} here, which is constructed using HST F814W observations. 

The inclination of this source is low, and thus errors in the inclination could introduce large uncertainties. Thanks to the well resolved CO disc in this object, however, we can constrain the inclination well and do not expect this to significantly change this result. This is shown graphically in Figure \ref{imffig}, where all error bars on the points have already been marginalised over our inclination uncertainty.

This object has a fairly uniform old stellar population which is predicted to have a high mass-to-light ratio ($\approx$5 in $i$-band assuming a Salpeter IMF), which increases slightly towards the galaxy centre. However, our CO observations require a much lower M/L, which is also modestly centrally peaked. Thus despite being a massive ETG, the IMF in this object seems to be uniformly light, approximately Kroupa like. 

\subsubsection{NGC3607}

NGC3607 is a fast-rotating ETG with a central dust disc that is part of the Leo II Group of galaxies. The molecular disc is regular and relaxed at our resolution, with no sign of the disturbed structures present in the dust distribution. Its surface brightness profile is well fitted with an exponential disc model.

The stellar population M/L of this (fast-rotating) elliptical has a mild negative gradient, which is consistent with that predicted dynamically from our CO observations. The IMF mismatch parameter in this source seems thus to be flat with radius, with a Kroupa/Chabrier type normalisation.

 \subsubsection{NGC3665}

 NGC3665 is an isolated fast-rotating ETG with a strong AGN, which powers large radio jets \citep{1986A&AS...64..135P,2016MNRAS.458.2221N}. The molecular gas and dust in this source is concentrated in a central disc perpendicular to the radio jet. As discussed in {Onishi et al., in prep} the best fitting surface brightness profile for this object is an exponential disc with a central hole, which we adopt here. We also adopt the mass model used in that study, constructed using $H$-band NICMOS observations. 
 
 This object has a fairly uniform stellar population. The dynamically inferred M/L, on the other hand, shows an M/L dip at around $\approx$350pc. This is approximately where the CO distribution peaks in this object so we could be seeing the dilution of the old stellar population with new stars at this location, however, one would also then expect to detect a decrease in the stellar population M/L. Overall the IMF mismatch parameter seems to lie midway between Kroupa and Salpeter in this object.  
 
\subsubsection{NGC4429}

NGC4429 is a lenticular galaxy in the Virgo cluster with a nuclear disc of gas and dust. The molecular disc is regular and relaxed, with a small hole in its centre (radius $\approx$0\farc45, 36pc; {Onishi et al., in prep}). This object has a slight M/L gradient over the extent of the molecular gas disc, with an $r$-band M/L of $\approx$6. Our CO measurements, however, reveal that the central regions of this object contain more mass than expected, and the M/L varies between $\approx$7.5 to 6.5 within the inner kiloparsec. This is also reflected in the IMF mismatch parameter, which is marginally consistent with Salpeter in the outer parts but heavier in the galaxy centre.

\subsubsection{NGC4459}
 
 NGC4459 is a lenticular galaxy in the Virgo cluster with a nuclear disc of gas and dust, which is discussed in detail above. Analysis of its stellar population suggests some radial M/L variation, with a gentle rise in the M/L with radius, and a M/L dip around the 200pc ring/spiral features. This M/L feature could arise because the relative amount of recent star formation in this region is higher, as the molecular gas is concentrated at this radius after streaming inwards from the spiral spurs. Our stellar population analysis does detect a  younger population in the inner parts of the this object, but the low mass fraction of this component does need lead to a significant change in the predicted M/L. This feature could also be due to the non-circular inflowing motions, as discussed below. Overall, however, despite these features the IMF appears to be consistent with (or slightly heavier than) Kroupa in this object.

\subsubsection{NGC4526}
 
NGC4526 is a lenticular galaxy in the Virgo cluster with multiple rings/spirals of molecular gas in its centre (see \citealt{2013Natur.494..328D,2015ApJ...803...16U}). 
The stellar population M/L of this galaxy is flat with radius within the molecular rich region, which is consistent with the flat gradient predicted dynamically from our CO observations.  A small peak in the dynamically inferred M/L is present around 200pc, consistent with the location of the largest CO ring. Apart from this small feature the IMF mismatch parameter in this source seems to be flat with radius, with a Salpeter-type normalisation.

  \subsubsection{IC0719}
 \label{discussIC0719}
 
 IC0719 is a relatively low mass lenticular galaxy (M$_{*} = 4.3\times10^{10}$ M$_{\odot}$; \citealt{2013MNRAS.432.1709C}), with strong star formation all across its disc triggered by a recent accretion of molecular gas \citep{2011MNRAS.417..882D}. The gas exactly counter-rotates when compared with the stars, and some minor flux asymmetry is still present. Despite this, the gas appears kinematically relaxed, and sits in the galaxy mid-plane, and thus is is well suited to our modelling approach. This object is quite young (SSP equivalent age of 3.4Gyr within 1 Re, \citealt{2015MNRAS.448.3484M}), and has a large velocity dispersion gradient, making it an interesting case study for IMF variation studies. 
 
We make use of 3.6$\mu$m observations from the S4G survey (\citealt{2010PASP..122.1397S}) to derive the luminous stellar mass profile of this galaxy. Non-stellar 3.6$\mu$m emission has been subtracted from this image, using the procedure defined in \cite{2012ApJ...744...17M}. Using this band allows us to minimise the effect of dust obscuration, as the source is very dusty even in the SDSS $z$-band, and 2MASS near-infrared observations are not deep enough. The molecular gas resolution is worse than the Spitzer PSF at these wavelengths, and thus we do not expect the low resolution of the 3.6$\mu$m observations to affect our results. 
 
A strong dynamical M/L gradient seems to be needed to explain the observed rotation of the gas, as shown in the right panel in Figure \ref{imffig}. A 3.6$\mu$m M/L of $\approx$0.2 is required in the central bin (coloured grey here), while beyond the inner 500pc the 3.6$\mu$m M/L becomes $\approx$1.0. The M/L estimated from our stellar population fits at 3.6$\mu$m is much more constant, with values around unity. This leads us to predict to a strong $\alpha_{\rm dyn}$ gradient in this source, with the IMF normalisation increasing outwards, but being consistent with Salpeter beyond the innermost 500pc.

{Note that although our line-strength based metallicity measurements suggest a light weighted metallicity above the cutoff of -0.4 in the MIUSCAT-IR models, the optical star-formation history carries weight down to [M/H]=-1.71 at the level of 5-10\% in the centre, to 50\% in the very outer parts (>10 arcsec). As mentioned above, this could lead to an underestimation of the population M/L by 10-20\% \citep{2014ApJ...788..144M}, which would make our final IMF lighter, more consistent with Kroupa/Chabrier. We add an additional 20\% error onto each M/L$_{\rm pop.salp}$ estimate here to take this systematic into account.}
 
 {Although the CO emission in this object is well fit by an exponential disc model (as shown in the top left hand panel of Figure \ref{imffig}) this source is fairly edge on and so a central hole in the CO emitting disc cannot be ruled out.  A hole is visible in the distribution of H$\beta$ flux in our SAURON maps, however we do not know if the distribution of molecular gas has the same feature. If present, such a hole could lead to a spurious decrease of the inner dynamical M/L. An exponential disc model without a hole is preferred by our MCMC modelling. However, higher resolution observations will be required to determine the presence/absence of a central hole entirely.} 
   
 {In addition, if present, hot dust emission from the torus around an active galactic nucleus (AGN) in this object could affect the innermost M/L measurement. This object is classified as a broad-line AGN in the Sloan Digital Sky Survey \cite{2000AJ....120.1579Y}, but evidence for nuclear activity was not found in our \atlas\ observations \citep[e.g.][]{2016MNRAS.458.2221N}. In any case, given the uncertainty in this innermost measurement we do not consider it in the rest of this work. Including/removing this point does not alter any of our conclusions.}

\subsection{Comparison with previous works}
\label{sec:imfcomparison}

     \begin{figure}
\begin{center}
\includegraphics[width=0.475\textwidth,angle=0,clip,trim=0cm 0cm 0cm 0.0cm]{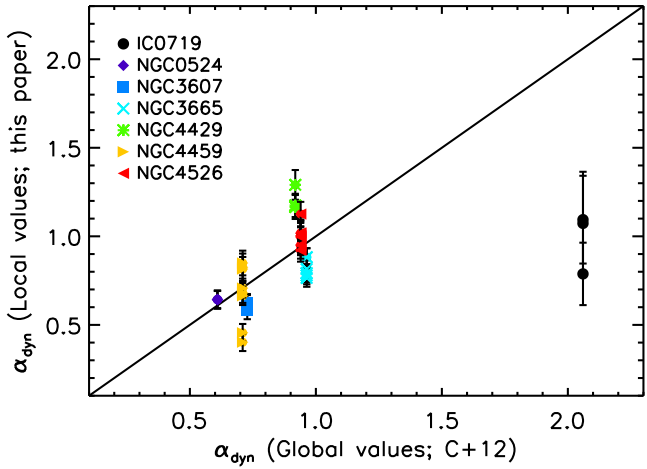}
 \caption{Comparison of the global IMF mismatch factor found by \protect \cite{2012Natur.484..485C} with the local values derived in this work. The radial bins within each source are shown in the same colour and symbol, as indicated in the legend. A one-to-one correlation is shown with the black line to guide the eye.}
 \label{fig:imfcomparison}
 \end{center}
 \end{figure}

Given that this is the first time that IMF parameters have been derived using molecular gas, it is important to compare our results to those found by other authors. As our targets are draw from the \atlas\ survey they are included in the study of \cite{2012Natur.484..485C}, who derive their own IMF mismatch parameter for each source. They assume each source has a single applicable M/L, but given the relatively flat gradients we see here it is not unreasonable to compare their value to that found in each of our radial bins. This comparison is shown in Figure \ref{fig:imfcomparison}. The radial bins within each source are shown in the same colour, as indicated in the legend. 

In general we find good agreement between our radially resolved approach and that unresolved study, with our values differing by $\ltsimeq$0.1 dex from those of \cite{2012Natur.484..485C}. The exceptions are NGC4429 (where our $\alpha_{\rm dyn}$ is slightly higher than that of \citealt{2012Natur.484..485C}), and IC0719, where our $\alpha_{\rm dyn}$ is radically different.
It should be noted, however, that IC0719 was explicitly excluded from much of the analysis in \cite{2012Natur.484..485C,2013MNRAS.432.1709C} because of the strong gradient in age within the object which makes unresolved analyses (where M/L gradients are assumed to be absent) unreliable. Our resolved analysis should be more reliable in such a case.

 We hence conclude that overall our study is in good agreement with the dynamical studies of \cite{2012Natur.484..485C} and \cite{2013MNRAS.432.1709C}, as expected given the analysis conducted in \citealt{2013MNRAS.429..534D} comparing CO and stellar kinematics.
 
Two of our objects were included in the work of \cite{2012ApJ...760...71C}, NGC0524 and NGC4459. That work used gravity sensitive stellar features to spectroscopically derive the IMF normalisation, and as such is interesting to compare with our dynamical method. NGC0524 was found in \cite{2012ApJ...760...71C} to have an IMF normalisation similar to the Milky Way ($\alpha_{\rm pop,MW}$=1.08). When transformed to our system with a Salpeter IMF as the reference this equates to $\alpha_{\rm pop,salp}$=0.7. Our value averaging over all radii is consistent with this ($<\alpha_{\rm dyn}>$=0.66). NGC4459 in our work has a Kroupa like IMF normalisation ($\alpha_{\rm dyn}$=0.70), while \cite{2012ApJ...760...71C} report a light IMF $\alpha_{\rm pop,salp}$ of 0.52 ($\alpha_{\rm pop,MW}$ of 0.8). This difference is significant on a $\approx2\sigma$ level. The disagreement in this object highlights the point made by \cite{2014MNRAS.443L..69S}, that further work must be done in this field to compare techniques. Overall in this work, however, with only two comparison objects it is hard to draw general conclusions.

\begin{figure}
\begin{center}
\includegraphics[width=0.475\textwidth,angle=0,clip,trim=0cm 0cm 0cm 0.0cm]{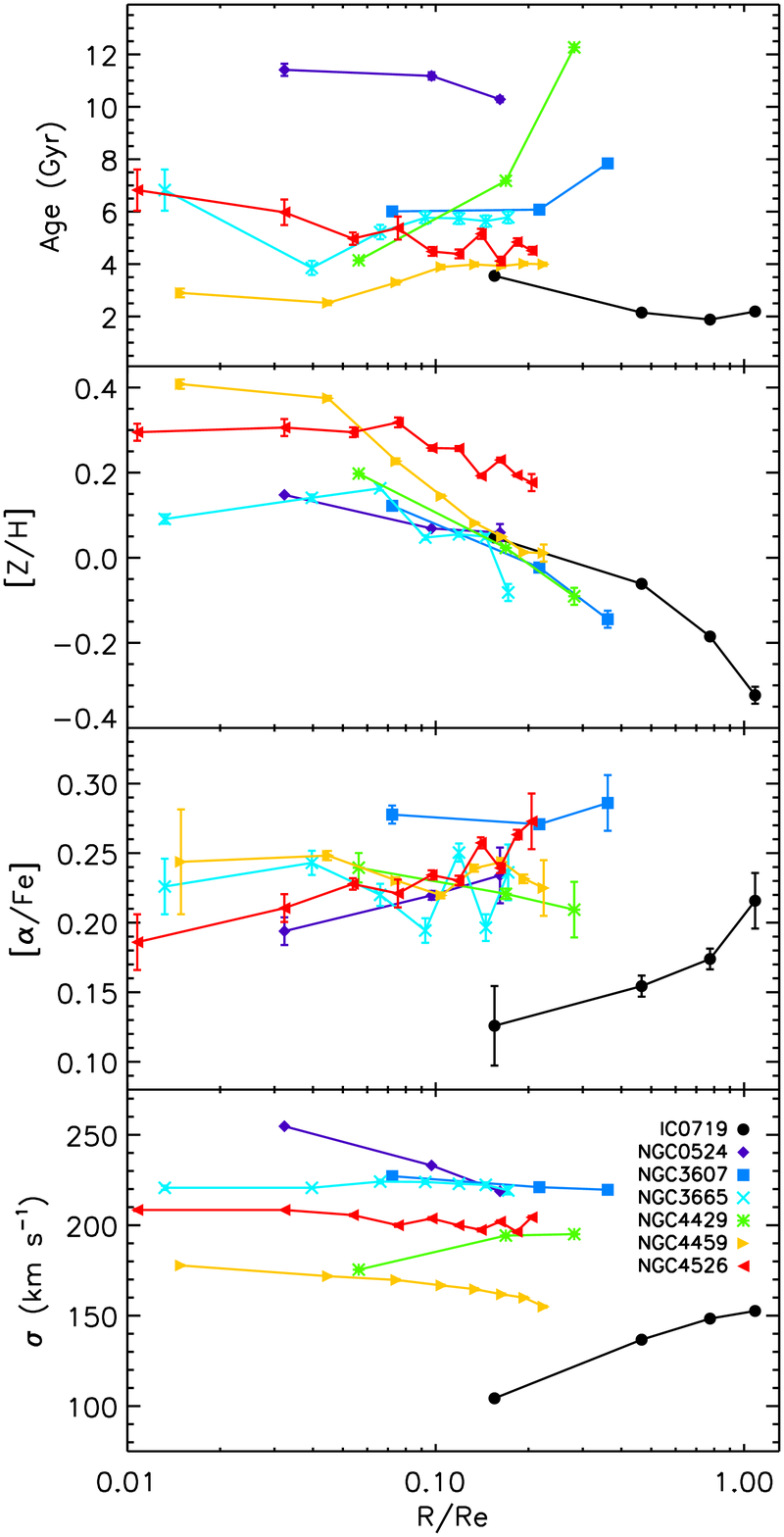}
\caption{Radial variation of resolved SSP equivalent stellar population parameters measured within our elliptical beam-width bins. We show the weighted mean age, metallicity and alpha-enhancement of the stars, in the first, second and third panel, respectively. Also shown in the fourth panel is the radial variation of the velocity dispersion.}
 \label{fig:radial_galprops}
 \end{center}
 \end{figure}

   \begin{figure}
\begin{center}
\includegraphics[width=0.475\textwidth,angle=0,clip,trim=0cm 0cm 0cm 0.0cm]{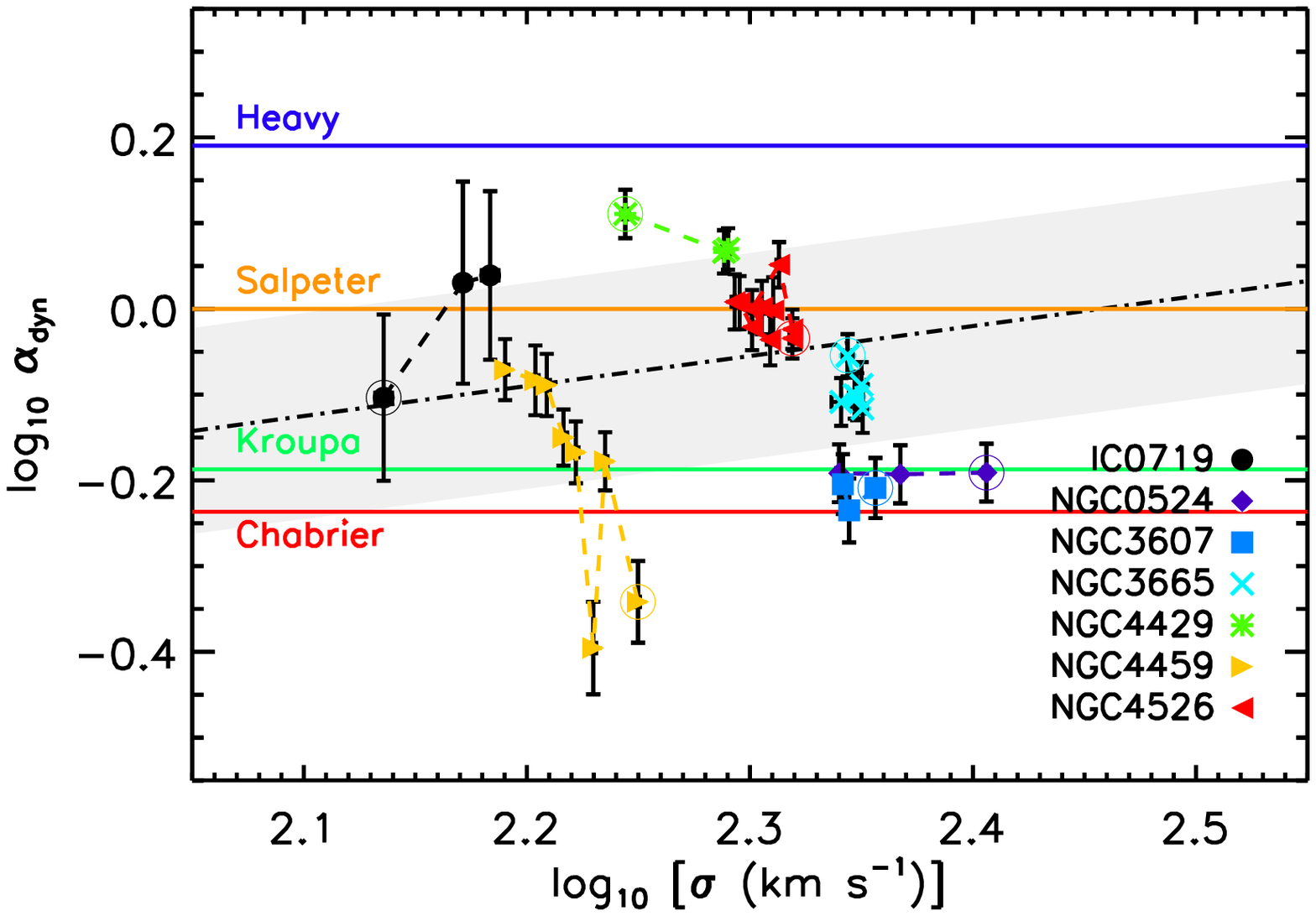}\vspace{0.25cm}
\includegraphics[width=0.475\textwidth,angle=0,clip,trim=0cm 0cm 0cm 0.0cm]{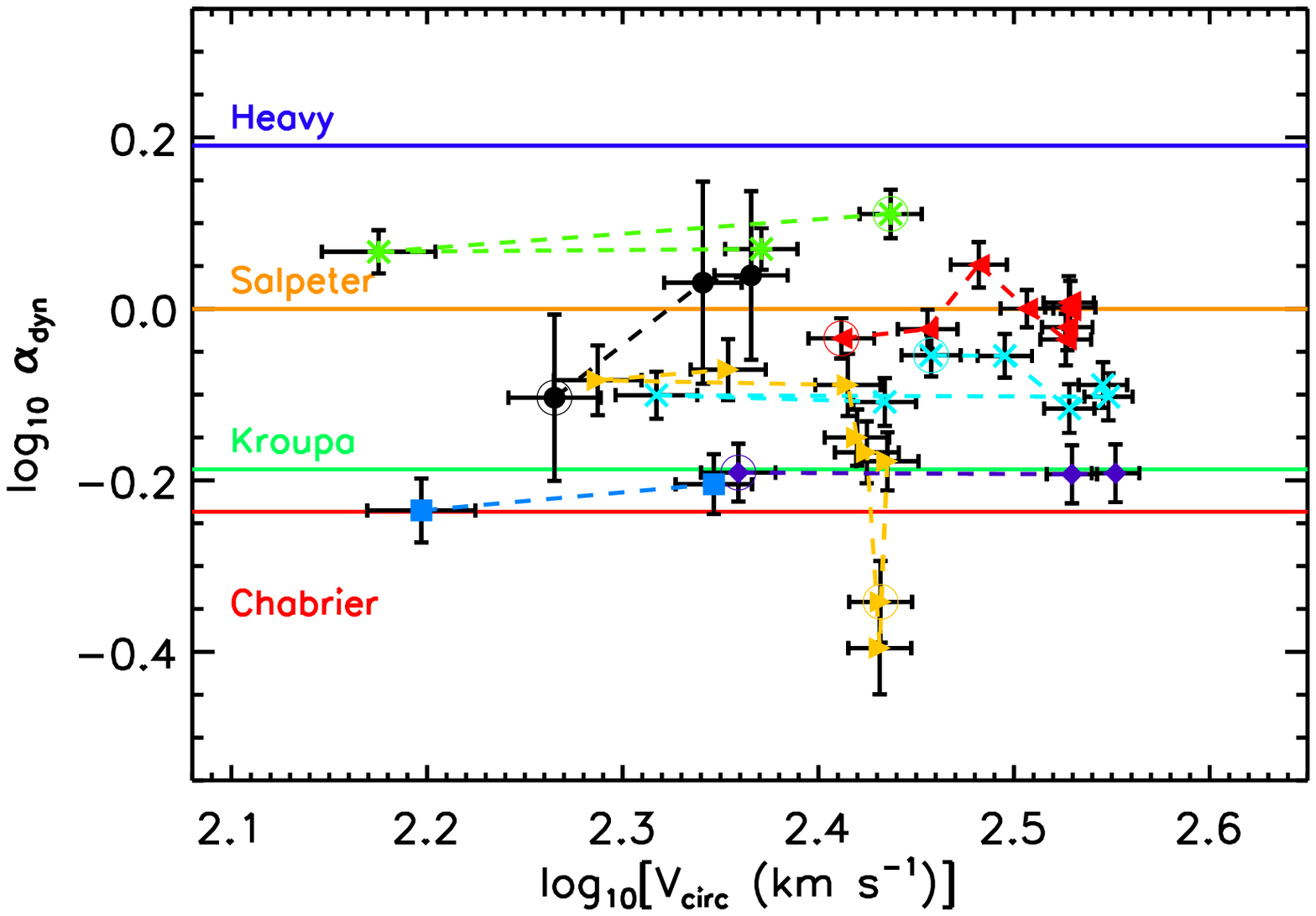}
 \caption{The base-10 logarithm of the IMF mismatch parameter ($\alpha_{\rm dyn}$), plotted against resolved dynamical properties of the host galaxy. \textit{Top panel:} log of the mean stellar velocity dispersion within each bin. 
 \textit{Bottom panel:} the log of the circular velocity (measured at the central radius of the bin). In both plots individual radial bins within each source are shown in the same colour and symbol, as denoted in the legend. The radial bins are joined with a dashed line of the appropriate colour, and the innermost bin we consider is highlighted with a larger open circle in addition to the normal symbol, in order to allow readers to discern radial trends. Also shown are solid coloured lines that denote the $\alpha_{\rm dyn}$ parameter of a heavy, Salpeter, Kroupa and Chabrier IMF. A black dot-dashed line is shown in the top panel, denoting the best fit correlation of \protect \cite{2012Natur.484..485C,2013MNRAS.432.1709C}, and the grey bar around it denotes the reported scatter around that relation of 0.12 dex.}
 \label{fig:imfsigma}
 \end{center}
 \end{figure}

  \subsection{The need for IMF variation}
  
Our objects show that no single IMF can match the observations. For instance in objects like NGC0524, NGC3607 and NGC4459 the enclosed mass implied if the stellar population had a Salpeter IMF is greater than the total dynamical mass. As this would be physically impossible, these objects clearly need a lighter IMF.  On the other hand, objects such as NGC3665, NGC4429 and NGC4526 seem to need a heavier IMF to explain their gas kinematics.

  \subsection{Variation with galaxy parameters}
\label{galparamvar}

In the sections above we have shown evidence that the IMF varies both within and between massive early-type galaxies, with a range of IMF mismatch parameters being found in our sample of seven molecular rich ETGs. In this section we attempt to determine if the IMF mismatch parameter correlates with any local properties of the the galaxy at those locations. 

We here utilise the \atlas\ IFU data to derive the average stellar velocity dispersion, circular velocity, and stellar populations  parameters (age, metallicity, [$\alpha$/Fe]) found in the IFU bins that fall within each beam-width radial region. These radial bins are elliptical (with an axial ratio defined by the galaxy inclination), and are the same as those used to derive the dynamical IMF constraints. The weighted mean of these variables within each annulus is shown as the point, with the error on the weighted mean as horizontal error bar on each point in the following figures. We use these estimates to explore putative correlations between these parameters and the IMF, as suggested by other authors \citep[e.g.][]{2012ApJ...760...71C,2012Natur.484..485C,2013MNRAS.433.3017L,2015ApJ...806L..31M}.

{We begin by showing the radial variation of each of these key parameters in Figure \ref{fig:radial_galprops}. Our objects tend to have fairly flat alpha-enhancement and velocity dispersion gradients within the region probed, negative metallicity gradients, and a dichotomy of age profiles. In the rest of this section we show how these radial variation compare with the $\alpha_{\rm dyn}$ values we derive above.}

\begin{figure}
\begin{center}
\includegraphics[width=0.475\textwidth,angle=0,clip,trim=0cm 0cm 0cm 0.0cm]{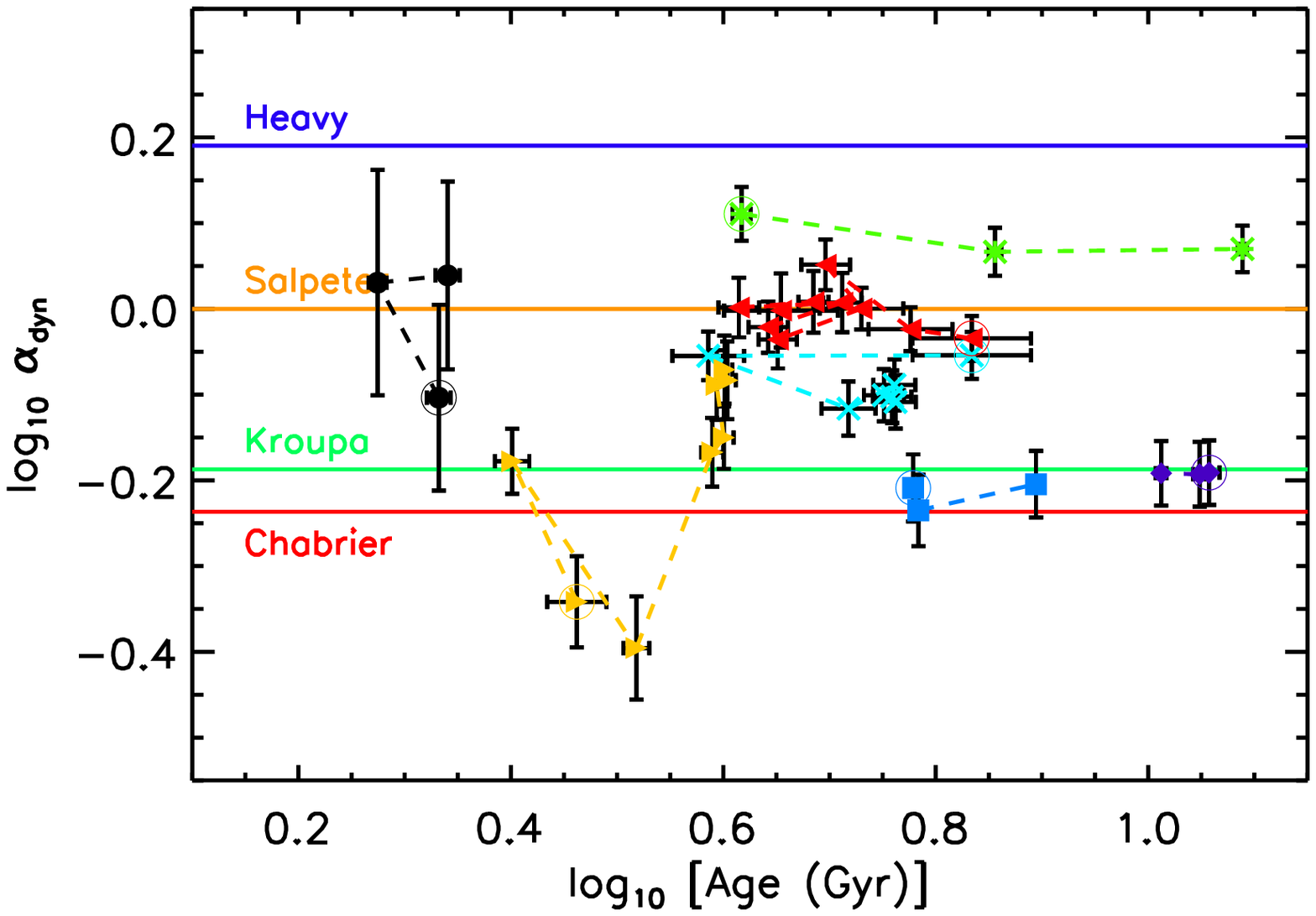}\vspace{0.25cm}
\includegraphics[width=0.475\textwidth,angle=0,clip,trim=0cm 0cm 0cm 0.0cm]{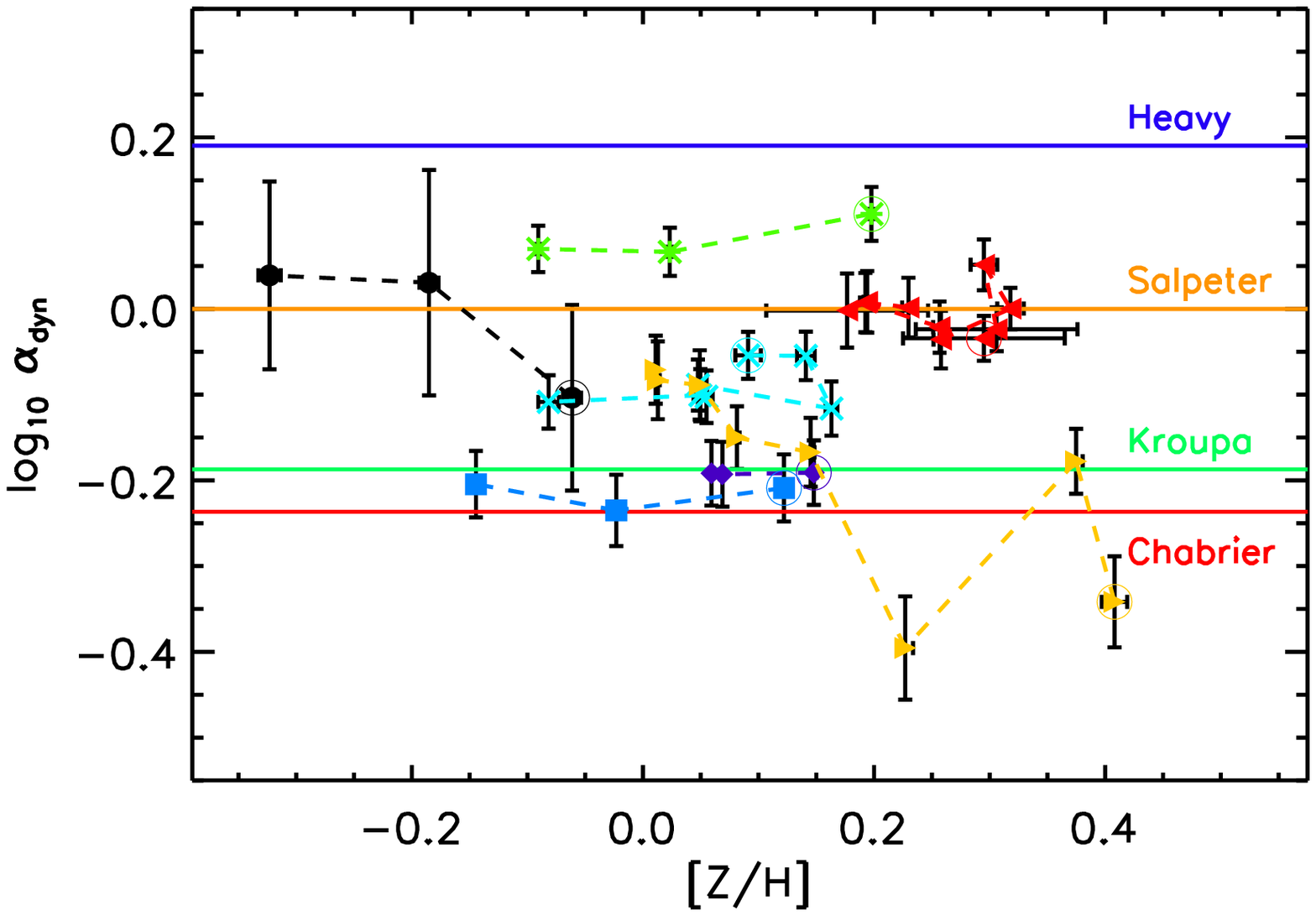}\vspace{0.25cm}
\includegraphics[width=0.475\textwidth,angle=0,clip,trim=0cm 0cm 0cm 0.0cm]{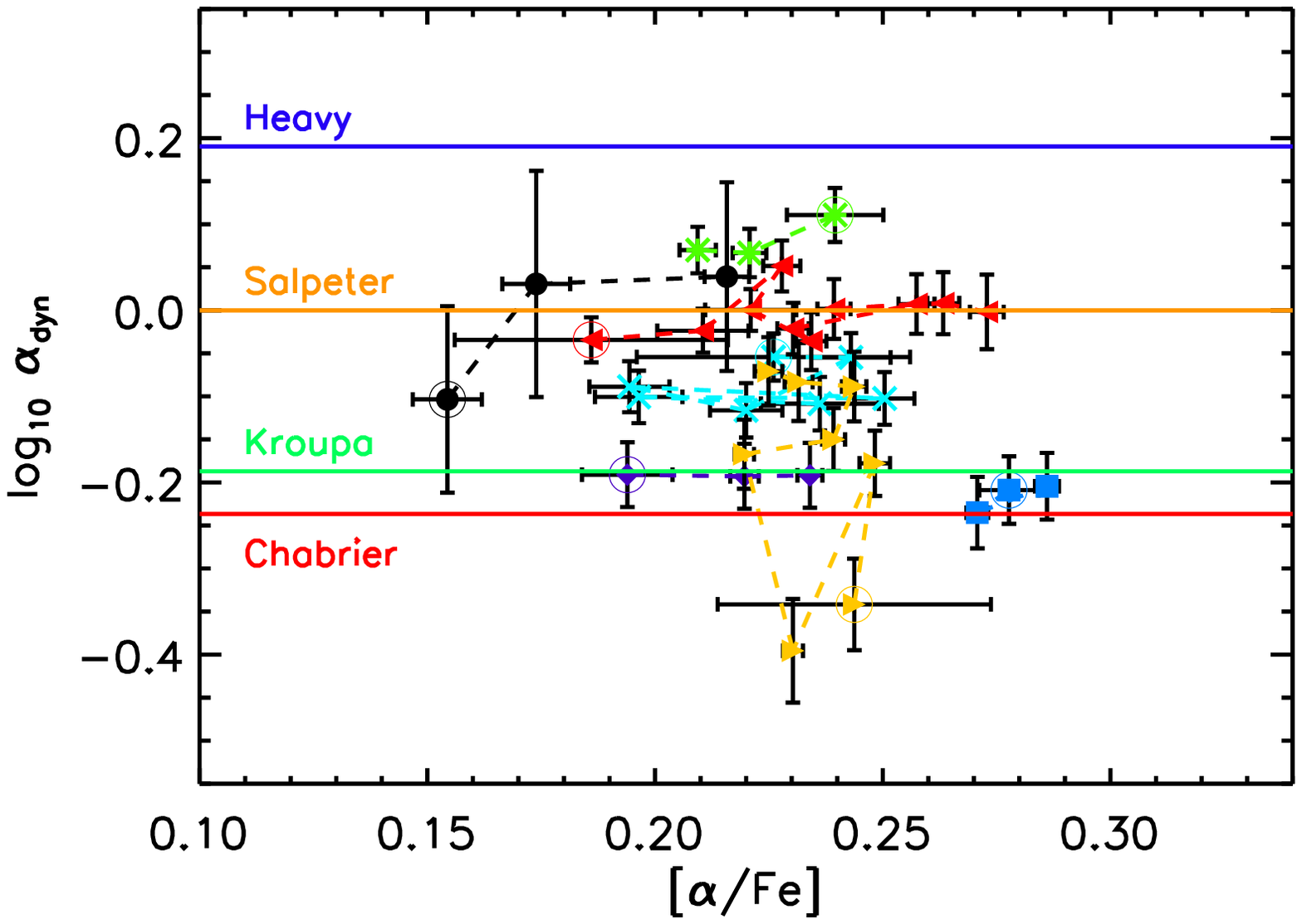}
\caption{As Figure \protect \ref{fig:imfsigma}, but showing the logarithm of the IMF mismatch parameter ($\alpha_{\rm dyn}$), plotted against resolved SSP equivalent stellar population parameters within that bin. These are the mean age, metallicity and alpha-enhancement of the stars, in the top, middle and bottom panel, respectively.}
 \label{fig:imfstellpop}
 \end{center}
 \end{figure}

  \subsubsection{IMF variation with dynamical properties}

The top panel of Figure \ref{fig:imfsigma} shows the IMF mismatch parameter ($\alpha_{\rm dyn}$) found within each radial bin of our sources, plotted against the mean stellar velocity dispersion within that bin\footnote{Available from www.purl.org/atlas3d} \citep{2011MNRAS.413..813C}.  Also shown are coloured lines that denote the $\alpha_{\rm dyn}$ parameter of a heavy, Salpeter, Kroupa and Chabrier IMF. No clear and consistent correlation is seen in this figure, either between objects, or radially within individual galaxies. A black dot-dashed line is shown denoting the best fit correlation of \protect \cite{2012Natur.484..485C}, with grey shading showing its scatter.

The bottom panel of Figure \ref{fig:imfsigma} shows the IMF mismatch parameter ($\alpha_{\rm dyn}$) found within each radial bin of our sources, plotted against the circular velocity (measured from our MGE models at the central radius of the bin). The circular velocity gives a measure of the depth of the potential, and any systematic variation of the IMF with this parameter could imply that the potential of the host galaxy plays a role in determining the IMF. No clear correlation is seen in this figure, however, either between objects, or radially within individual galaxies. This is discussed further in Section \ref{discuss}.

     \subsubsection{IMF variation with stellar population parameters}

Figure \ref{fig:imfstellpop} shows the IMF mismatch parameter ($\alpha_{\rm dyn}$) found within each radial bin of our sources, plotted against the weighted mean SSP equivalent age, metallicity and alpha-enhancement of the stars within that bin (in the top, centre and bottom panels respectively).   As above labelled coloured lines denote the $\alpha_{\rm dyn}$ parameter of a heavy, Salpeter, Kroupa and Chabrier IMF. 

No clear correlation is observed between $\alpha_{\rm dyn}$ and these parameters, either locally or globally.
These findings are consistent with the weak or absent global correlations reported in \cite{2014ApJ...792L..37M} using the stellar dynamics based $\alpha_{\rm dyn}$ values from \cite{2012Natur.484..485C}.
We discuss further all the results on the variation of the IMF with galaxy parameters in Section \ref{discuss}.

\subsection{Main uncertainties}
\label{uncertainties}
In this Section we consider the main uncertainties that could affect the results derived following the method above. These can roughly be split into two subsets; issues with the modelling process, and breakdown of the physical assumptions we make to simplify complicated astrophysical problems.

\subsubsection{Dust and its effect on luminous mass models}

In order to convert the velocity profiles estimated from our dynamical modelling to dynamical mass-to-light ratio measurements, as described above we here parameterise the luminous matter distribution using MGE models of the stellar light distribution.  However, it is possible that remaining dust contamination would cause us to underestimate the luminosity of the stars in some parts of our object, and thus bias the derived dynamical M/L. The analysis conducted here is especially affected by this form of uncertainty, as these objects have been selected to host a cold ISM. In order to minimise this issue our MGE models were constructed from images at the longest wavelength available, in order to minimise dust contamination. The models are also carefully fitted to remove any contamination from dust still visible in the image (via masking of affected regions). Such a process has been shown to work well at recovering the intrinsic light distribution in such systems \citep{2002MNRAS.333..400C}. Some underlying uncertainty remains, which could cause our dynamical M/L's to be biased to high values. The objects with the most dust obscuration usually have low predicted IMF normalisations (Kroupa/Chabrier like), however, suggesting that this uncertainty cannot remove the need for a heavy IMF in some objects.

\subsubsection{Non-circular motions and gas velocity dispersion}
In this analysis we assume that the gas is purely in circular rotation. If significant non-circular motions (e.g. inflow, outflow, streaming motions) exist within their gas reservoirs then this could affect our analysis.

{\cite{2015arXiv150904881R} studied the effect of non-circular motions on the derivation of mass profiles in detail, and showed dramatic variations can be caused in strongly barred galaxies if the bar is orientated in specific directions with respect to our line of sight. In order to avoid this problem we have purposefully selected sample galaxies with regular, relaxed, dynamically cold discs (see \citealt{2013MNRAS.429..534D}) in galaxies that are not strongly barred (or the gas is within the inner Lindblad resonance of the bar where gas inflowing motions are minimal; \citealt{1985A&A...150..327C,1989Natur.338...45S}).}

{Non-circular motions can still be present in objects without bars, for instance because of other departures from axisymmetry in the gravitational potential (i.e., spiral arms). Such motions are low in amplitude compared to the rotation of these systems (for example $\approx$10 \kms\ in M51, a galaxy with much stronger spiral structure than these early-types; \citealt{2013ApJ...779...45M}) , and add uncertainty at the $\approx$5-10\% level.}
Such motions also exist only at specific locations within a molecular gas disc, and so the smooth gradients found in our objects argue that non-circular motions are not a major effect.
The kinematic fits we present here are generally good (with low reduced $\chi^2$ values of 1 -- 3), suggesting we do not have significant non-circular motions present within the discs that our models cannot fit. 

As discussed above, one object (NGC4459) does show some signatures of non-circular motions. We suspect it has low level inflowing motions around its innermost ring (see above), that are likely the cause of a slightly lower dynamical M/L in this region. This difference is small, however, and does not significantly skew the IMF estimate of this source as a whole.

{In addition to non-circular motions, an additional source of uncertainty comes from the molecular gas velocity dispersion. A simple prescription for dealing with this is included in our models, which allow for a single characteristic velocity dispersion in each disc. Our modelling procedure find values in the range of 5-15 \kms, typical for nearby galaxies \citep{2013AJ....146..150C}. The uncertainty in this measurement has already been marginalised over when producing the error bars on the M/L and $\alpha_{\rm dyn}$ measurements.  In reality, however, the velocity dispersion could change with radius, and position within the gas disc. If this increase occurs within the flat part of the galaxy rotation curve, away from the central regions of the galaxy then no bias would be expected in our dynamical estimates, as this would simply broaden the line-width of the gas symmetrically around the same best fit value. In the central part of the galaxy where beam smearing is important, however, an increase in velocity dispersion could cause a slight overestimation of M/L$_{\rm dyn}$ (see e.g. \citealt{2016ApJ...823...51B}).}

{In order to quantify the size of the effect a variable gas velocity dispersion could have on our derived parameters we re-ran the modelling process for NGC4459, allowing the velocity dispersion to vary independently in each of the beam-width sized radial regions we fit. NGC4459 was chosen for this test, as it is the object in our sample most likely to suffer from these problems (as argued above). While some variation of the velocity dispersion was preferred by the fit (variations of $<$5 \kms), this had very little effect on the derived M/L, causing a mean absolute variation in the derived M/L$_{\rm dyn}$ values of less than 3\%, which is small compared to the other errors. }

\subsubsection{Dark matter and gas mass}

IMF analyses are generally affected by the assumptions they make for the distribution of dark matter. It can be very hard to separate the affect of extra mass caused by a heavy IMF, from that caused by a different dark matter distribution. In this work we are helped, however, by only being able to probe the central regions of early-type galaxies. Various studies of the dark matter content in ETGs have showed that these inner regions are totally dominated by the mass of stars. For instance, \cite{2013MNRAS.432.1709C} find typical dark matter fractions of $13\%$ within the effective radius of ETGs. As our observations typically probe only within a quarter of the effective radius, we expect the dark matter contribution to the mass profile to be very low.

\begin{figure}
\begin{center}
\includegraphics[width=0.475\textwidth,angle=0,clip,trim=0cm 0cm 0cm 0.0cm]{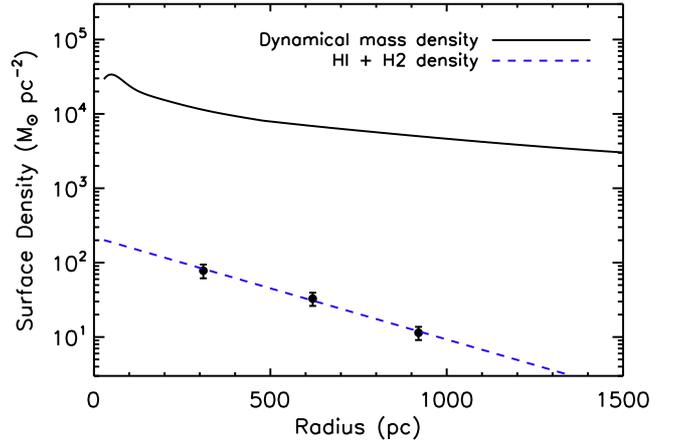}
\caption{Mass surface densities in NGC0524. The dynamical mass density (black line) is derived from the observed circular velocity curve. The molecular and atomic gas density points are from \protect \cite{2013MNRAS.432.1914M}, while the blue line is the best fitting exponential disc model. The cold gas contributes at most a hundredth of the mass density at any radius.}
 \label{massproffig}
 \end{center}
 \end{figure}

In addition to dark matter, the mass of any ISM material present also contributes to the total dynamical mass of the system. The hot phases of the ISM are diffuse enough to not contribute significantly to the mass budget in the central parts of galaxies, but both \hi\ and H$_2$ masses could potentially affect the mass profiles we derive. This has been shown to be an important affect in gas-rich objects, especially at high redshifts \citep[e.g.][]{2014A&A...565A..59D}. It should be possible to include the gas mass in our calculations, and thus take it into account. However, given the uncertainty in $X_{\rm CO}$ (the conversion between CO luminosity and mass) this could induce large errors. In the centre of early-type galaxies, however, the density of the ISM material is insignificant compared to the stellar mass density (for any sensible $X_{\rm CO}$). This is shown graphically in Figure \ref{massproffig}, which shows the dynamical mass surface density and the cold gas surface density in NGC0524, which was previous studied in detail by \cite{2011MNRAS.410.1197C} and \cite{2013MNRAS.432.1914M}. 
The dynamical mass surface density (assuming the mass is spherically distributed) is derived from the rotation curve as

\begin{equation}
\Sigma_{\rm dyn} = \frac{V^2}{\pi G r} ,
\end{equation}
where $V$ is the circular velocity derived from our MGE and M/L model fitting, $G$ is the gravitational constant and $r$ is the radius. This is compared to the atomic and molecular gas surface density (calculated assuming the mass is distributed in a flat disc), which is taken from the observations of \cite{2013MNRAS.432.1914M} (in this object the molecular gas density dominates over the atomic gas). The cold gas contributes at most 1\% of the mass density at any radius, making it dynamically unimportant. We thus consider it unlikely that neglecting the mass of gas in our fitting procedure biases our results. 

\section{Discussion}
 \label{discuss}

 In the Sections above we presented our method to measure dynamical mass-to-light ratios as a function of radius in seven ETGs. We compared these measurements to stellar population parameters derived from IFU data in order to estimate the IMF-mismatch parameter, and how it varies with radius. In this Section we discuss these results, and what they may mean in a wider context. 
 
 \subsection{Radial M/L variation in molecular gas rich ETGs}

In the right panel of Figure \ref{imffig} we show the mass-to-light ratio gradients present within our seven target ETGs. To first order, ETGs such as these have uniformly old stellar populations with a frosting of new stars formed in the most recent gas-rich merger/accretion episode (see e.g. \citealt{2014MNRAS.444.3408Y}). These young stars have a very low mass-fraction, but may contribute significant amounts of light, especially in the blue/UV. This is likely to lower the M/L of the stellar population within the molecular gas disc.

In a disc of star-forming gas the new stars will be formed following the spatial distribution of the gas. Unless the galaxy light also follows an exponential disc profile we thus expect mass-to-light gradients to exist within the disc. 
Several of our objects do show small M/L gradients (both in M/L$_{\rm pop}$ and M/L$_{\rm CO}$), and several other objects have smaller variations at individual radii, which relate to specific features in their gas distribution (see the discussion of individual sources above). Overall, however, our analysis shows that M/L gradients do not seem to be very strong in the inner parts of these objects (whichever method you use to derive them). This is likely because of the low star formation rates of these sources. The objects that do seem to have some M/L variations are also those with the highest specific star formation rates \citep{2014MNRAS.444.3427D}. In the other objects even luminous O/B stars cannot outshine the more numerous old stars in the cores of these objects (especially in the long wavelength bands we use to create our luminous mass models).

  \subsection{IMF variation, and correlations with galaxy parameters}
  
  Above we showed that no single IMF can match the observations of gas kinematics in our sample ETGs, and thus that the IMF appears to be variable. With this technique we are only sensitive to the total IMF normalisation, and are unable to determine if the extra mass comes from high or low mass stars. Other analyses suggest the latter to be more likely, but we cannot comment further on this here.

In Section \ref{galparamvar} we showed that the IMF normalisation derived in radial bins in our sample ETGs does not seem to correlate with various galaxy parameters. This is true both locally within individual galaxies, and globally between systems. This is true for both dynamical (galaxy circular velocity or stellar velocity dispersion; see Figure \ref{fig:imfsigma}) and stellar population parameters (age, metallicity and alpha-enhancement). 

The galaxy parameters we consider here all show significant radial variation in galaxies. This highlights the importance of understanding aperture effects, which complicate the comparison of the IMF derived from (spatially weighted) kinematics with luminosity-weighted stellar populations analyses. This study allows us to remove potential issues with aperture effects that were faced by previous dynamical studies, by considering the local variations in both IMF and stellar populations. We still find no evidence for a correlation between IMF and stellar populations, consistent with the weak or absent correlations reported in \cite{2014ApJ...792L..37M}. 

The lack of correlation with stellar velocity dispersion may seem surprising when one remembers the good agreement between the IMF parameters derived in this study and those from \cite{2012Natur.484..485C} (as shown in Section \ref{sec:imfcomparison}), who report a strong correlation between these variables. The best fit reported by \cite{2012Natur.484..485C} is shown in Figure \ref{fig:imfsigma}. Our objects do not seem to individually follow this relation, but as a whole the majority of the points fall within $\approx$0.2 dex of the correlation. The 1$\sigma$ scatter around the best fit in \cite{2012Natur.484..485C} was 0.12 dex (shown as the grey shaded region in Figure \ref{fig:imfsigma}), and as $\approx$2/3 of our points fall within this region, we cannot rule out the possibility that our objects global normalisations have been drawn from this underlying distribution. If this is the case, however, then any correlation with stellar velocity dispersion appears to be global, rather than also appearing locally within galaxies.

In this work we find no significant correlations with any of the studied dynamical or population parameters, and thus it seems that substantial disagreements remain between different studies of IMF variation. For instance the resolved stellar population study of \cite{2015ApJ...806L..31M} found a strong correlation between the IMF normalisation and metallicity, and the work of  \cite{2012ApJ...760...71C} suggested the strongest correlation was between the IMF and alpha-element enhancement.
Our analysis does seem to yield reasonably consistent $\alpha_{\rm dyn}$ values to those found in other works (e.g \atlas; see Figure \ref{fig:imfcomparison}), however small differences are found in one of the two objects that overlap with the study of \cite{2012ApJ...760...71C}. Future studies of a greater number of objects that will help reveal if these studies still lack internal consistency \citep{2014MNRAS.443L..69S}. Given that our results are consistent with those using stellar kinematics, however, it does suggest that if a problem exists in current analyses it is likely to lie in the uncertainties inherent in stellar population modelling, rather than our understanding of galaxy dynamics.

\section{Conclusions}
\label{conclude}

In this paper we present the first spatially-resolved study of the IMF in external galaxies derived using a dynamical tracer of the mass-to-light ratio. We do this using molecular gas kinematics in seven early-type galaxies selected from the \atlas\ survey. We compare these measurements to stellar population parameters derived from star formation histories in order to estimate the IMF-mismatch parameter, and shed light on the variation of the IMF within early-type galaxies. 

We find that the mass-to-light ratio gradients in the inner parts of our target objects are not very strong (independent of the method used to derive them). Several objects do show modest M/L$_{\rm CO}$ gradients, while other have smaller M/L$_{\rm CO}$ variations at individual radii, which relate to specific features in their gas distribution. The objects that do seem to show M/L$_{\rm CO}$ variations are also those with the highest specific star formation rates. The majority of these slight gradients are also present in the stellar population analyses, but not all are. 

We confirm that the IMF appears to vary when comparing different massive ETGs. Some of our target objects require a light IMF, otherwise their stellar population masses would be greater than their dynamical masses. In contrast, other systems seem to require heavier IMFs to explain their gas kinematics. We find good agreement between our IMF normalisations derived using molecular gas kinematics and those derived by \atlas\ using stellar kinematics. This provides an independent check on the stellar kinematic results, suggesting that if a problem exists in current analyses it is more likely to lie in the stellar population modelling, rather than in the dynamics. We note that this agreement occurs despite our objects being the hardest to model using stellar methods (due to their strong discs of gas and dust). Thus with this technique we can, in principle, extend studies of the IMF normalisation to more gas-rich systems. 

We do not see strong variation of the IMF normalisation with galaxy dynamical or stellar population properties in this work, either locally or globally. 
This study allows us to remove potential biases due to aperture effects that were faced by previous studies. By considering the local variations in both IMF and stellar populations we find no evidence for a correlation between IMF and stellar populations, consistent with the weak or absent correlations reported in \cite{2014ApJ...792L..37M}. In this work alone we also do not find a convincing connection between galaxy dynamics and the IMF. Future works of this type will be required to show if the lack of observed correlations is real, or due to low number statistics. 

In the future larger studies of molecular gas kinematics can help to disentangle the cause of IMF variation. This method provides an independent check on the conclusions of other dynamical methods, and in addition can provide dynamical information on IMF gradients within individual galaxies. Projects such as the CARMA-EDGE survey \citep{2015IAUGA..2257914U} and the MASSIVE survey \citep{2014ApJ...795..158M}, which combine IFU kinematics with resolved molecular gas observations should allow extension of this technique to large samples of galaxies across the Hubble sequence. 

 \vspace{0.5cm}
\noindent \textbf{Acknowledgments}

TAD acknowledges support from a Science and Technology Facilities Council Ernest Rutherford Fellowship, and thanks F. van de Voort for fruitful discussions. 

This paper is based on observations carried out with the IRAM Thirty Meter Telescope, the IRAM Plateau de Bure interferometer and the CARMA interferometer. IRAM is supported by INSU/CNRS (France), MPG (Germany) and IGN (Spain). Support for CARMA construction was derived from the states of California, Illinois, and Maryland, the James S. McDonnell Foundation, the Gordon and Betty Moore Foundation, the Kenneth T. and Eileen L. Norris Foundation, the University of Chicago, the Associates of the California Institute of Technology, and the National Science Foundation.

This paper is based in part on observations made with the NASA/ESA Hubble Space Telescope, and obtained from the Hubble Legacy Archive, which is a collaboration between the Space Telescope Science Institute (STScI/NASA), the Space Telescope European Coordinating Facility (ST-ECF/ESA) and the Canadian Astronomy Data Centre (CADC/NRC/CSA). This research has made use of the NASA/IPAC Extragalactic Database (NED) which is operated by the Jet Propulsion Laboratory, California Institute of Technology, under contract with the National Aeronautics and Space Administration.

\bsp
\bibliographystyle{mnras}
\bibliography{bibIMF.bib}
\bibdata{bibIMF.bib}
\bibstyle{mnras}

\label{lastpage}

\end{document}